\documentclass[twocolumn]{aastex631}

\usepackage [english]{babel}
\usepackage [autostyle, english = american]{csquotes}
\MakeOuterQuote{"}

\usepackage{amsmath}

\begin{document}

\title{
Saving Doomed Planets: \\ Mass Loss and Angular Momentum Return Boost Hot Jupiter Survival Rates}

\correspondingauthor{Grant C. Weldon}
\email{gweldon@astro.ucla.edu}

\author{Grant C. Weldon}
\affiliation{Department of Physics and Astronomy, UCLA, Los Angeles, CA 90095, USA}
\affiliation{Mani L. Bhaumik Institute for Theoretical Physics, Department of Physics and Astronomy, UCLA, Los Angeles, CA 90095, USA}

\author{Bradley M. S. Hansen}
\affiliation{Department of Physics and Astronomy, UCLA, Los Angeles, CA 90095, USA}
\affiliation{Mani L. Bhaumik Institute for Theoretical Physics, Department of Physics and Astronomy, UCLA, Los Angeles, CA 90095, USA}

\author{Smadar Naoz}
\affiliation{Department of Physics and Astronomy, UCLA, Los Angeles, CA 90095, USA}
\affiliation{Mani L. Bhaumik Institute for Theoretical Physics, Department of Physics and Astronomy, UCLA, Los Angeles, CA 90095, USA}

\begin{abstract}
The existence of giant extrasolar planets on short-period orbits ("hot Jupiters") challenges planet formation theories because such planets are difficult to form close to the star. High-eccentricity migration is a leading explanation, in which giant planets born at large separations are excited to near-unity eccentricities, enabling tidal dissipation at periastron to shrink and circularize their orbits. While observations of orbital misalignments and eccentric planets support this scenario, high-eccentricity migration models struggle to reproduce the observed hot Jupiter occurrence rate. Population synthesis studies often predict that many source "cold Jupiters" are destroyed by tidal disruption at high eccentricities. We revisit this question with improved treatments of mass loss and angular momentum return experienced by tidally perturbed planets. As a test case, we explore eccentricity excitations driven by wide stellar companions via the Eccentric Kozai-Lidov (EKL) mechanism. We show using an analytical framework that planets may avoid complete disruption and ultimately survive as stripped hot Jupiters. To capture detailed planetary mass loss over many orbits, we perform numerical studies that combine secular dynamical evolution with planetary structure evolution. Our new population synthesis studies show that hot Jupiter survival is enhanced by a factor of $\sim2-3$ relative to previous estimates, yielding occurrence rates ($\gtrsim 0.5\%$ around FGK stars) consistent with observations. Angular momentum return from mass accreted onto the star may also produce a pileup of hot Jupiters near three-day orbital periods. These results suggest that high-eccentricity migration, when accounting for tidal mass loss, may be a dominant channel for hot Jupiter formation.
\end{abstract}
\keywords{Exoplanets, planetary systems, exoplanet dynamics}

\section{Introduction} \label{sec:intro}

Since the discovery of the first exoplanet around a Sun-like star in 1995 \citep[][]{Mayor+95}, the presence of giant planets on close-in orbits ($a<0.1$ au) has surprised astronomers \citep[for a review, see][]{Dawson+18}. These extreme ``hot Jupiters'' have no analogue in our Solar System, where Jupiter and Saturn reside at several au from the Sun.

The core accretion model of giant planet formation posits that a $\gtrsim 10 M_{\oplus}$ rocky core must be built to accrete a giant gaseous envelope \citep[e.g.,][]{Pollack+96}. Although models for \textit{in situ} formation of hot Jupiters do exist \citep[e.g.,][]{Batygin+16,Boley+16}, most models start with the formation of a gas giant on scales $\gtrsim1$ au, where more solid material is available to form the rocky cores. The planets must then migrate inward to their present locations. Migration may be driven by a gas disk \citep[e.g.,][]{Goldreich+80,Lin+86}, or from tidal dissipation at high eccentricities induced by planet-planet scattering  \citep[e.g.,][]{Rasio+96,Ford+05,Zhou+07,Juric+08,Chatterjee+08,Nagasawa+08,Ford+08,Nagasawa+11,Carrera+19}, secular chaos in multi-planet systems \citep[e.g.,][]{Wu+11,Teyssandier+19}, or secular perturbations from a faraway planetary or stellar companion \citep[e.g.,][]{Holman+97,Wu+03,Takeda+05,Fabrycky+07,Naoz+11,Naoz+12,Lithwick+11,Katz+11,Teyssandier+13,Petrovich+15a,Petrovich15b,Anderson+16,Stephan+18,Stephan+20,OConnor+21,Stephan21,Stephan+24,Weldon+24,Weldon+25}. 

High-eccentricity tidal migration has garnered support due to the preference for hot Jupiters to have planetary or stellar companions capable of inducing high eccentricities \citep[e.g.,][]{Knutson+14,Ngo+15,Ngo+16,Zink+23}, the elevated eccentricities of potential source, cold Jupiters \citep[e.g.,][]{Weldon+25}, and the presence of giant planets on intermediate, eccentric orbits that suggest ongoing migration \citep[e.g.,][]{Naef+01,Dong+21,Gupta+24}. Despite growing support for high-eccentricity migration, there remains a problem with the underlying theory. Population synthesis studies have found that this mechanism cannot produce nearly enough hot Jupiters to match the observed hot/cold ratio of $\sim10-15\%$ \citep[e.g.,][]{Naoz+12,Petrovich+15a,Petrovich15b,Munoz+16,Anderson+16,Dawson+18}. The issue lies not in the inefficiency of generating high eccentricities via scattering or secular interactions, but rather the high rate ($\sim10-40$\%) at which cold planets are lost to tidal disruption during eccentric passages in which the pericenter crosses the Roche limit. These prior studies assumed a simplified treatment in which planets are always destroyed upon crossing the Roche limit. In this work, we use an improved treatment for tidal mass loss to revisit the question of whether planets are always completely disrupted or whether some planets may only experience partial disruption during close passages and ultimately survive as hot Jupiters. Recovering only a small fraction of the large population of lost planets would significantly boost the theoretical production rate, potentially alleviating the long-standing tension between theory and observation. 

As a test case, we will focus on the high-eccentricity migration scenario in which a distant companion induces eccentricity oscillations via secular perturbations, known as the Eccentric Kozai-Lidov (EKL) mechanism \citep[e.g.,][]{Kozai62,Lidov62,Naoz16}. We will specifically consider the case where the distant perturber is a bound stellar companion to the hot Jupiter host. Indeed, hot Jupiters have been found to preferentially exist in wide stellar binaries \citep[e.g.,][]{Knutson+14,Ngo+15,Ngo+16,EelesNolle+25}. While recent work by \cite{Moe+21} did not find such a correlation, \cite{Stephan+24} suggested that many seemingly single hot Jupiters could have formed through EKL in binaries that were later unbound by white dwarf kicks. EKL also tends to produce misalignments between the host stellar spin axis and planetary orbit normal. Such misalignments are observed \citep[e.g.,][]{Triaud+10,Albrecht+12}, particularly around hot stars, where thinner convective zones are thought to be less efficient at producing tidal realignment \citep[][]{Winn+10}. Many warm Jupiters ($0.1$ au $<a<$ $0.8$ au)\footnote{The upper boundary of 0.8 au for warm Jupiters is defined in \cite{Weldon+25} as the average distance, for Sun-like stars, where general relativistic precession overwhelms EKL, allowing for tidal circularization.}---potentially in the process of tidal migration---also display misalignments in the presence of a stellar companion \citep[e.g.,][]{Rice+22,Gupta+24}. Therefore, the stellar EKL mechanism serves as a well-motivated test to determine whether improved treatment of tidal mass loss enhances the formation rate of hot Jupiters.

The model setup for the three-body systems, treatment of mass loss, and initial conditions are discussed in Section~\ref{sec:modelsetup}. In Section~\ref{sec:analytical}, we use an analytical framework to model the effects of tidal mass loss and conduct population synthesis studies. More detailed numerical methods are employed in Section~\ref{sec:numerical}, including planetary structure and orbit response to mass loss. We discuss implications of our work and future directions in Section~\ref{sec:discussion}. We summarize our conclusions in Section~\ref{sec:conclusion}.

\section{Model Setup}
\label{sec:modelsetup}

We adopt the model in which a gas giant initially orbits a Sun-like star at distances close to the snow line, where gas giant formation from core accretion is expected to occur \citep[e.g.,][]{Hayashi+81,Pollack+96}. The star+planet pair
is orbited by a distant stellar companion, in a hierarchical configuration. Our goal is to examine the efficiency with which such configurations excite the eccentricity
of the planetary orbit to order unity values. We then examine the range of outcomes for populations of systems. In this section, we discuss the aspects of the model that are common to both the analytical and numerical approaches that we will take. 

\label{sec:methods}
\subsection{Hierarchical three-body system}

The inner star+planet system (stellar mass $m_*$ and planet mass $m_p$) with semi-major axis $a_p$, is orbited by a companion (in this work, a star) with mass $m_c$ and semi-major axis $a_c$. We denote the angle of inclination of the inner (outer) orbit with respect to the total angular momentum by $i_1$ ($i_2$), and the mutual inclination between the two orbits is $i$ = $i_1 + i_2$. 

For hierarchical systems, the three-body Hamiltonian can be averaged over the orbital periods and expanded in powers of the small semi-major axis ratio $\alpha = a_p/a_c$ \citep[e.g.,][]{Kozai62, Harrington68, Ford+00, Naoz+13}. At the quadrupole level (proportional to $\alpha^2$), the inner orbit can undergo oscillations of eccentricity and inclination on timescales much longer than the orbital periods \citep{Kozai62, Lidov62}. The quadrupole level of approximation is often insufficient when the outer companion's orbit is eccentric or when the planet mass is non-negligible \citep{Naoz+13}. In these cases, the octupole contribution (proportional to  $\alpha^3$) can drive the inner orbit to extremely high eccentricities or even flip the orbit from prograde to retrograde with respect to the total angular momentum \citep[see for a full set of equations,][]{Naoz16}. 

\subsection{Tidal mass loss}
\label{sec:tidalmassloss}

Prior studies of the secular production of hot Jupiters assumed planets are completely tidally disrupted immediately when the periastron $q$ crosses some multiple of the Roche limit $r_t$ \citep[e.g.,][]{Naoz+12,Petrovich15b,Anderson+16}, defined as
\begin{equation} \label{eq:roche}
    r_t = R_p \left(\frac{m_*+m_p}{m_p}\right)^{1/3} \ ,
\end{equation}
where $q =2.7r_t$ is typically chosen as a strict boundary for complete disruption. This value is informed by \cite{Guillochon+11}, in which 3-D hydrodynamical simulations of multi-orbit close encounters between eccentric giant planets and host stars were performed. This work found that planets begin losing mass near $q \sim 2.7r_t$, and recent work on the Roche limit modified by dynamical tides found a similar threshold where mass loss may begin \citep[][]{Yu+25}. Over several orbits, \cite{Guillochon+11} found that planets with periastra $q <2.7r_t$ are completely disrupted. Planets are only partially disrupted during the first close passage, but interactions between the surviving planet and tidally-stripped debris heat the planet, leading to increased mass loss on subsequent passages. These interactions also act to destabilize the orbit and can eject the planet from the system. 

There is reason to believe that mass loss occurs more gently for secularly-driven giant planets than in \cite{Guillochon+11}, due to fewer and more gentle interactions between surviving planets and stripped debris. The assumptions used in the \cite{Guillochon+11} simulations following the planetary passage do not necessarily apply to the systems that we are interested in. \cite{Guillochon+11} considered a system on an unperturbed Keplerian orbit. In our model, emerging tidal streams will be subject to precession induced by the distant perturber. Any difference in planetary and streamer orbital parameters will drive differential precession, reducing the degree of interaction at periapse. For the same periapse distance, the cold Jupiters in this work tend to occupy more eccentric ($e_p\sim 0.99$) orbits than those in the hydrodynamical simulations ($e_p=0.9$), requiring a larger semi-major axis that allows for additional separation between the planet and debris streams\footnote{The EKL precession rate scales as $\dot{\omega_1} \propto a_p^3$ \citep[e.g.,][]{Naoz16}, and the time to re-encounter material on an orbit as $T \propto a_p^{3/2}$ \citep[e.g.,][]{Kepler1619}.}.

An additional concern is that the fluid in the hydrodynamic simulations is modeled as an $n=1$ polytrope. This approximation is appropriate for the dense, partially degenerate interior of a giant planet, but once the mass is liberated from the planet's gravity, it should expand adiabatically \citep[e.g.,][]{Spitzer39}. Planetary encounters with diffuse material should be significantly less violent than with high density material. Additional effects, such as stellar winds, may also drive the lost mass away from the planetary orbit \citep[e.g.,][]{Spalding+20}. Altogether, these effects make it likely that a partially disrupted planet's orbital evolution is less influenced by encounters with tidal debris streams on subsequent passages.

In our models, we consider planets with periastra $q<2.7r_t$ to be stripped planets that are not necessarily fully disrupted. In the analytical framework, we treat these systems as a population of planets with a fixed amount of mass loss. In the more detailed numerical runs, we examine the mass loss on an orbit-by-orbit basis using the results of \cite{Guillochon+11} and assuming little re-accretion of material onto the planet. Once mass is lost from the planet, it may ultimately accrete onto the star and return angular momentum to the planet (``conservative mass transfer''), or it may leave the system and carry the lost angular momentum away (``non-conservative mass transfer''). It is also possible that some fraction of the mass goes into both channels. In this work, we explore each of these three possible scenarios by considering angular momentum return fractions $f_{\rm ret}$ of 0, 0.5, and 1. In traditional tidal disruption event studies, it is generally taken that half of the stripped material accretes onto the central object \citep[e.g.,][]{Rees+1988}. Therefore, we consider $f_{\rm ret}=0.5$ to be the fiducial value.

\subsection{Initial conditions of population}
\label{sec:initialconditions}

To facilitate comparison with earlier work, we adopt initial conditions consistent with those used in previous studies, following the setup of \cite{Weldon+25}, which reliably reproduces the results of other authors employing similar assumptions. Building on this foundation allows us to isolate and assess the impact of the new physical models introduced in this study.

Systems are drawn using a Monte Carlo approach. The stellar masses of the host ($m_*$) and companion ($m_c$) are drawn between 0.6–1.6 $M_{\odot}$ (roughly FGK stars) from a Salpeter initial mass function with $dn/dm \propto m^{-2.35}$ \citep{Salpeter55}. Planet masses $m_p$ are drawn between 0.3–10 $M_J$ from a power-law distribution with $dn/dm_p \propto m_p^{-1}$ \citep{Marcy+2000}. 

Orbital parameters are sampled as follows. Planetary semi-major axes $a_p$ are drawn uniformly between 0.5–6 au. Outer-orbit periods are drawn from the log-normal distribution of \citet{Duquennoy+91}, restricted to $a_2 = 50$–1500 au. For simplicity in the analytical model, we assume $e_p = 0$ initially. In the numerical simulations, planetary orbit eccentricities $e_p$ follow a Rayleigh distribution with mean 0.13, consistent with small perturbations from planet–planet or planet–disk interactions \citep[as discussed in e.g.,][]{Weldon+25}. Outer-orbit eccentricities $e_c$ are drawn uniformly between 0–1 \citep[e.g.,][]{Raghavan+10,Moe+17}\footnote{Some works adopt a thermal distribution \citep[e.g.,][]{Jeans19}, which favors higher eccentricities and thus stronger EKL effects. Our uniform choice provides a more conservative estimate.}. The mutual inclination is drawn isotropically (uniform in $\cos i$ from $0$–$180^\circ$). Arguments of periapse $\omega_1$ and $\omega_2$ are sampled uniformly from $0$–$360^\circ$.

All of our initial conditions are stable according to the \cite{Mardling+01} stability criterion
\begin{equation}
    \frac{a_c}{a_p}>2.8\left(1+\frac{m_c}{m_*+m_p}\right)^{2 / 5} \frac{\left(1+e_c\right)^{2 / 5}}{\left(1-e_c\right)^{6 / 5}}\left(1-\frac{0.3 i}{180^{\circ}}\right) \ .
\end{equation}
The systems are also sufficiently hierarchical for stability. That is, the hierarchy parameter
\begin{equation}
\label{eq:eps}
    \epsilon = \frac{a_p}{a_c} \frac{e_c}{1-e_c^2} \ ,
\end{equation}
fulfills $\epsilon<0.1$ \citep[e.g.,][]{Naoz16}.

\section{Analytical model}
\label{sec:analytical}

We first present an analytical framework to study the effects of partial tidal mass loss and angular momentum return on the population of hot Jupiters. The analytical model we use is a modified version of the framework in \cite{Munoz+16}. In their work, they calculate the maximum eccentricity induced via the EKL mechanism, accounting for the limiting effects of general relativity and tides. Then, this maximum eccentricity may be compared with the critical periastron required for tidal migration or tidal disruption to determine the system's outcome. Using this model, we perform a Monte Carlo population synthesis to understand the implications for hot Jupiter demographics.

We begin by summarizing the model from \cite{Munoz+16}. Under the assumption that the planet has zero initial eccentricity, at the quadrupole level of approximation, $e_{\rm max}$ satisfies \citep[][]{Munoz+16}
\begin{equation}
\label{eq:emax_quad}
\begin{gathered}\epsilon_{\mathrm{GR}}\left(\frac{1}{j_{\min }}-1\right)+\frac{\epsilon_{\text {Tide }}}{15}\left(\frac{1+3 e_{\max }^2+\frac{3}{8} e_{\max }^4}{j_{\min }^9}-1\right) \\ =\frac{9}{8} \frac{e_{\max }^2}{j_{\min }^2}\left(j_{\min }^2-\frac{5}{3} \cos ^2 i\right)\end{gathered} \,
\end{equation}
where $j_{\rm min} \equiv \sqrt{1-e_{\rm max}^2}$. The dimensionless quantities 
\begin{eqnarray}
    \epsilon_{\rm GR} &= &2.96 \times 10^{-2} \left(\frac{m_*}{M_{\odot}} \right)^2 \left(\frac{m_c}{M_{\odot}}\right)^{-1} \left(\frac{a_c}{100 \, \rm{au}}\right)^3 \left(\frac{a_p}{1 \, \rm au} \right)^{-4} \nonumber \\ &\times &(1-e_c^2)^{3/2} \ ,
\end{eqnarray}
and 
\begin{align}
\epsilon_{\rm tide} = \, & 9 \times 10^{-8} 
   \left(\frac{m_*}{M_{\odot}} \right)^2 
   \left(\frac{m_c}{M_{\odot}}\right)^{-1} 
   \left(\frac{m_p}{M_J}\right)^{-1} 
   \left(\frac{a_c}{100 \, \rm{au}}\right)^3 \notag \\
&
   \times \left(\frac{a_p}{1 \, \rm au} \right)^{-8}  \left(\frac{R_p}{R_J}\right)^5  
   \left( \frac{k_2}{0.25} \right) 
   (1-e_c^2)^{3/2} \ ,
\end{align}
represent the contributions to the precession from general relativity and tidal distortion, respectively. Here, $k_2$ is the Love number of the planet, which we take to be $k_2=0.25$\footnote{For reference, the Love number for Jupiter is approximately $k_2 = 0.57$ \citep[e.g.,][]{Durante+20} and for Saturn is approximately $k_2=0.32$ \citep[e.g.,][]{Kramm+11}. For two-layer models (core + fluid envelope), $k_2$ decreases with increasing ratio of core to total mass \citep[e.g.,][]{Kramm+11}. Therefore, we choose $k_2=0.25$ to reflect planets with massive cores, aligning with the MESA models in \ref{sec:MESA}.}.

The maximum value of $e_{\rm max}$ is reached when $i = 90^{\circ}$. This limiting eccentricity $e_{\rm lim}$  satisfies 
\begin{equation}
    \label{eq:emax_oct}
    \frac{\epsilon_{\rm GR}}{\sqrt{1-e_{\rm lim}^2}} + \frac{7}{24}\frac{\epsilon_{\rm tide}}{(1-e_{\rm lim}^2)^{9/2}} = \frac{9}{8} \ ,
\end{equation}
in the limit $e_{\rm lim}\to 1$. In the presence of a non-zero octupole potential, for sufficiently large inclinations, the eccentricity can be driven to very large values. The critical inclination $i_{\rm lim}$ beyond which this occurs depends on the strength of the octupole parameter $\epsilon$, see Eq.~(\ref{eq:eps}). \cite{Munoz+16} obtain a fitting formula for $i_{\rm lim}$ 
\begin{eqnarray}
    \cos^2i_{\rm lim} &= &0.26 \left(\frac{\epsilon}{0.1}\right) -0.536 \left(\frac{\epsilon}{0.1}\right)^2 +12.05 \left(\frac{\epsilon}{0.1}\right)^3 \nonumber  \\ &-&16.78 \left(\frac{\epsilon}{0.1}\right)^4 \ .
\end{eqnarray}
This relation holds for $\epsilon \leq 0.05$, and $i_{\rm lim} = 48^{\circ}$ for  $\epsilon > 0.05$. The maximum eccentricity for $i<i_{\rm \lim}$ (quadrupole case) is then approximated by Eq.~(\ref{eq:emax_quad}), and for $i \geq i_{\rm \lim}$ (octupole-active) is approximated by Eq.~(\ref{eq:emax_oct}). See  \cite{Lithwick+11} and \cite{Katz+11} for additional discussion on the criterion for the octupole contribution to become important under the test particle approximation. 

The critical periastron below which tidal dissipation is strong enough to circularize the planetary orbit and lead to hot Jupiter formation is 
\begin{align}
q_{\rm mig} = \, & 2.3 \times10^{-2}\, {\rm au} 
   \left( \frac{T}{10^8 \, \rm{yr}} \right)^{1/7} 
   \chi^{1/7} 
   \left(\frac{m_*}{M_{\odot}} \right)^{2/7} \notag \\
& \times \left(\frac{m_p}{M_J}\right)^{-1/7} 
   \left(\frac{R_p}{R_J}\right)^{5/7} 
   \left(\frac{a_p}{1 \, \rm au} \right)^{-1/7} \ ,
\end{align}
where $T$ is the time to circularize the planet and $\chi$ is a dimensionless tidal dissipation factor. We set $T = 10^9$ yr and $\chi=1000$ \footnote{$\chi = \Delta t_l /0.1 $s, where $\Delta t_l$ is the lag time. We take $\Delta t_l = 1/(2Qn)$, where $Q$ is the tidal quality factor and $n$ is the mean motion \citep[e.g.,][]{Eggleton98,Munoz+16}. When calculating $Q$, we use $t_v = 0.01$ yr to provide a direct comparison to the numerical runs. We note that $\chi$ is a highly uncertain quantity, and we have selected a highly dissipative, but characteristic value informed by calibrations by \cite{Petrovich15b}.}, though we note that the dependency of $q_{\rm mig}$ on these parameters is weak. For this analysis, we take a characteristic planetary radius of $R_p = 1.2 R_J$.

Planets are tidally disrupted for periastra below
\begin{equation}
    q_{\rm dis} = \, 4.8 \times 10^{-3} \, {\rm au}  \, \eta_{\rm dis} \left(\frac{R_p}{R_J}\right)  \left(\frac{m_*}{M_{\odot}} \right)^{1/3} \left(\frac{m_p}{M_J}\right)^{-1/3} \ .
\end{equation}
where $\eta_{\rm dis}$ historically is taken to be 2.7 \citep[from the completely disruptive multi-passage encounters in][as discussed in Section~\ref{sec:tidalmassloss}]{Guillochon+11}. In this analytical analysis, we set $\eta_{\rm dis} = 1.0$ to be the threshold for complete disruption.

Planets may lose a finite amount of mass and still survive as hot Jupiters. We consider the threshold for mass loss to be
\begin{equation}
    q_{\rm loss} = \, 4.8 \times 10^{-3} \, {\rm au}  \, \eta_{\rm loss} \left(\frac{R_p}{R_J}\right)  \left(\frac{m_*}{M_{\odot}} \right)^{1/3} \left(\frac{m_p}{M_J}\right)^{-1/3} \ .
\end{equation}
Here, we set $\eta_{\rm loss} = 2.7$, considering that planets may begin losing mass here, but only undergo partial disruption \citep[e.g.,][]{Guillochon+11,Yu+25}. Planets that obtain periastra $q_{\rm dis} < q < q_{\rm mig}$ are considered to be stripped hot Jupiters. 

This analytical model does not account for the detailed orbit-by-orbit response of the planet's structure and orbit to mass loss on individual passages, as we will explore in the numerical simulations. The primary purpose of this analytical exploration is to broadly understand the range of system outcomes as a function of the distribution of periastra from populations of perturbers.

\subsection{Rates of production}

\begin{figure*}[ht]
\begin{center}

\includegraphics[width=7.5in]{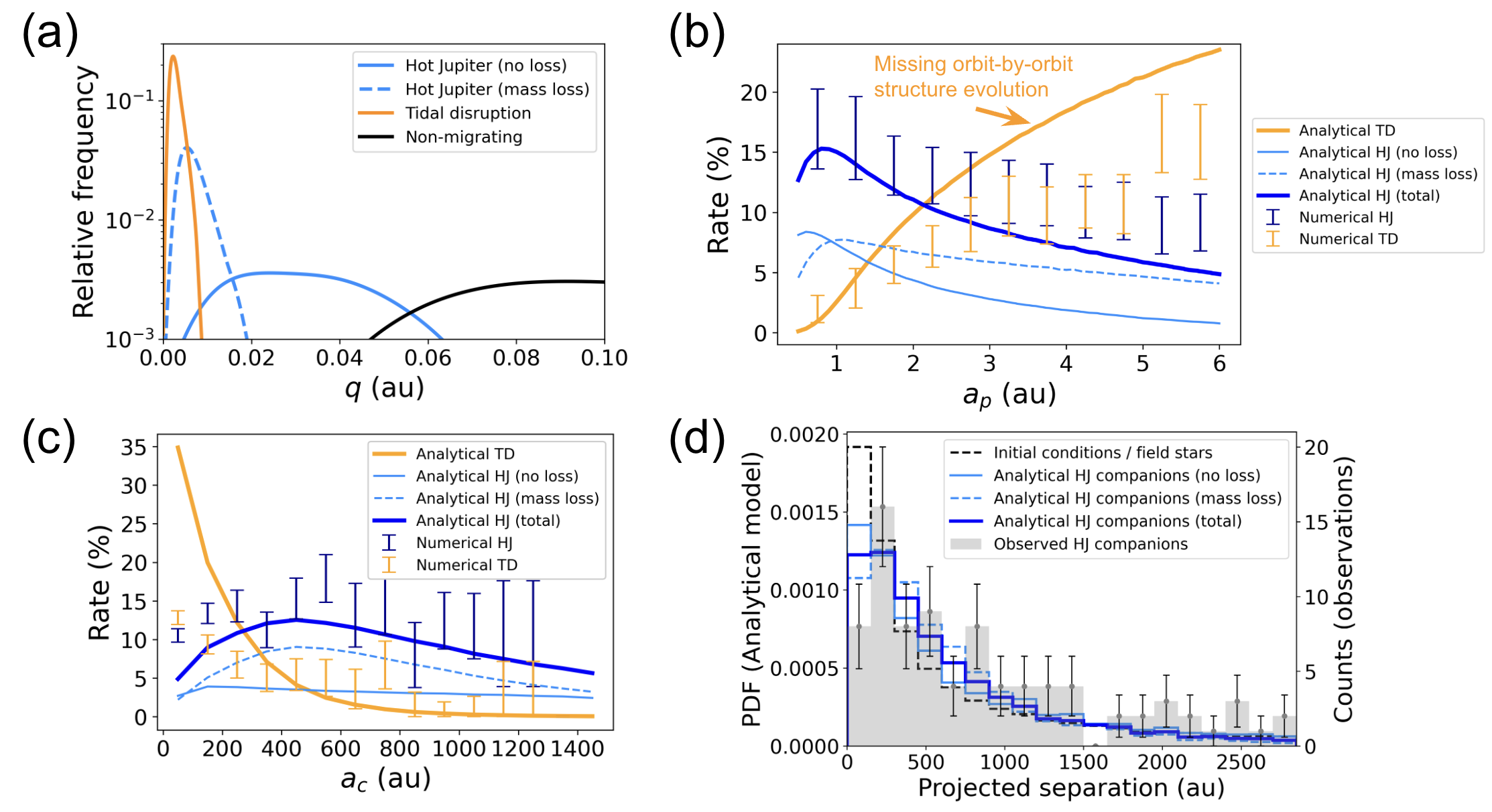}
\caption{\footnotesize Panel (a): Analytical distribution of periastra $q$ from the fiducial population of stellar perturbers, with color-coded outcomes determined by comparing to the critical $q$ for hot Jupiter formation (solid blue line), mass loss (dashed blue line), and tidal disruption (solid orange line). Planets that do not undergo migration or disruption are shown with a solid black line. The Monte Carlo results are smoothed using a Gaussian kernel. Panel (b): Analytical estimates for the rates of tidal disruption (solid orange line), hot Jupiter production with no mass loss (solid light blue line), hot Jupiter production with mass loss (dashed light blue line), and total hot Jupiter production (solid dark blue line). Data points show simulated results from the \texttt{M0} numerical run for comparison, where dark blue corresponds to hot Jupiters and orange to tidal disruptions. For each fixed planetary semi-major axis value $a_p$, a population of $10^6$ perturbers from $50<a_2<1500$ au is generated following \cite{Duquennoy+91}. Panel (c): Rates are color coded as in panel (b). For each stellar semi-major axis value $a_2$, a population of $10^6$ planets is drawn uniformly from $0.5<a_p<6$ au. Panel (d): Distributions of projected separations for field stars from \cite{Duquennoy+91} (dashed gray line), companions to hot Jupiters from the analytical model that do not undergo mass loss (solid light blue line), companions to hot Jupiters from the analytical model that undergo mass loss (dashed light blue line), companions to the overall analytical hot Jupiter population (solid dark blue line), and observed hot Jupiter companions (shaded gray histogram). The model distributions are displayed as a PDF and the observations in counts.}
\label{fig:analytic_model}
\end{center}
\end{figure*}

We first perform a Monte Carlo analysis using the initial condition distributions in Section~\ref{sec:initialconditions}. We note that in the numerical simulations, a Rayleigh distribution of initial planetary eccentricities is chosen, but the analytical model here assumes that planets start with zero eccentricity. Because initial eccentricities are relatively small, we expect the model to still provide a sufficient level of approximation, though it may slightly underestimate the number of hot Jupiters and tidal disruptions that are produced in the scenario where planets begin with non-zero eccentricities \citep[e.g.,][]{Weldon+25}. 

From a fiducial population of $10^6$ systems drawn from the distributions discussed in Section~\ref{sec:initialconditions}, we estimate the pericenter distance from Eq.~(\ref{eq:emax_quad}) or Eq.~(\ref{eq:emax_oct}), depending on whether the octupole is active. We then compare to $q_{\rm dis}$, $q_{\rm loss}$, and $q_{\rm mig}$. The distribution of $q$ and the various system outcomes at small periastra ($<0.1$ au) are shown in panel (a) of Figure \ref{fig:analytic_model}. For this fiducial population, 8.3\% of planets are tidally disrupted (solid orange line), 4.9\% of planets become hot Jupiters with no mass loss (solid blue line), 9.9\% of planets survive as stripped hot Jupiters (dashed blue line), and the remaining population does not migrate or become tidally disrupted (solid black line). The total hot Jupiter production rate is therefore 14.8\%, a factor of almost $3 \times$ higher than the rate obtained under the assumption of prior studies that all planets that undergo mass loss are completely disrupted.

Using the analytical model, we explore how the production rates depend on the initial location of the planet in panel (b) of Figure \ref{fig:analytic_model}. We vary the planet's initial location $a_p$ incrementally from 0.5 to 6 au in units of 0.1 au, and we draw a population of $10^6$ perturbers from $50<a_2<1500$ au. The rates of tidal disruption are shown with a solid orange line, the rates of hot Jupiter production with no mass loss as a solid light blue line, the rates of hot Jupiter production with mass loss as a dashed light blue line, and the summed rate of hot Jupiter production with a solid dark blue line. For comparison, the blue data points show the results of the \texttt{M0} numerical simulations for total hot Jupiters (discussed later in Section \ref{sec:numerical}), and the orange data points show numerical tidal disruptions from the same data set. 

We find a general agreement between the analytical model and the numerical simulations, considering that the simulations include orbit-by-orbit planetary structure evolution that may aid in the stability to mass loss. The rate of hot Jupiter production decreases slightly with larger $a_p$, as significantly more planets experience tidal disruption. However, the total rate of hot Jupiter production does not depend strongly on the initial choice of planetary location. Above $\sim1$ au, the dominant contribution to the hot Jupiter population switches from planets that undergo no loss to planets that do lose mass, owing to the stronger secular perturbations driving planets to lower $q$. We see overall in the analytical model that the new treatment for mass loss boosts the rate of hot Jupiter production by a factor of $\sim2-3\times$, an effect that is especially pronounced for larger $a_p$, where almost no hot Jupiters are expected to form under the traditional treatment due to the highly disruptive secular perturbations. 

In panel (c) of Figure \ref{fig:analytic_model}, we explore the dependence of hot Jupiter production on binary separation. We vary the perturber's separation incrementally from 50 to 1450 au in units of 100 au, and draw $10^6$ planets uniformly from 0.5 to 6 au in each bin. The rates are color coded in the same manner as in panel (b). Here, we note that the higher variance in the comparison numerical simulations is due to the non-uniform sampling of $a_c$ from the \cite{Duquennoy+91} distribution.

We see that the rate of tidal disruptions falls off strongly as $a_2$ increases, due to the weaker secular perturbations. Again, when accounting for mass loss, the hot Jupiter production rate is boosted by a factor of $\sim2-3$.  The rate of total hot Jupiter production rises out to $\sim500$ au before falling off. At separations below $\sim500$ au, the secular perturbations are sufficiently strong to drive a higher fraction of planets to periastra below the complete disruption threshold. At separations greater than $\sim500$ au, weaker perturbations allow many of these planets to survive as stripped hot Jupiters. Because of this effect, the hot Jupiter production rate is $\sim5-10\%$, regardless of perturber distance. Therefore, any underlying differences in the true binary separation distribution hosting the source population of cold Jupiters are unlikely to dramatically affect the hot Jupiter production rate.

In panel (d) of Figure \ref{fig:analytic_model}, we compare the distribution of companion separations in the analytical model to the observed distribution. We construct the observational sample by selecting hot Jupiters ($0.3M_J<m_p<10M_J$ and $a_p<0.1$ au) from the NASA Exoplanet Archive (accessed August 22, 2025) \citep[e.g.,][]{Akeson+13,PSCompPars} that are observed to have stellar companions and orbit host stars in the mass range $0.6M_\odot<m_*<1.6M_{\odot}$ (roughly FGK stars), although the stellar companion mass $m_c$ is not restricted to this range. For each each of the 102 represented systems, we take the projected separation of the companion from \cite{Schneider+11}, \cite{Mugrauer+19}, and \cite{Thebault+25}. The sample is incomplete and heterogeneous, though \cite{Thebault+25} argues that the large scale trends in the distribution of $a_c$ (e.g., lack of close-in planet-hosting binaries) cannot be explained by observation bias alone. Because many observed systems have M-dwarf companions, for comparison, we perform a new Monte Carlo analysis with the lower mass bound for $m_c$ extended to $0.1M_{\odot}$. Additionally, because the observations contain some very distant companions, we extend the maximum $a_c$ to 3000 au (a small number of systems have even higher $a_c$). All other parameters are sampled as previously described. To compare with observations, we project the analytical $a_c$ assuming random viewing angles.

There is agreement between the analytical and observed distributions of companion separations at the level of $\sim1-2\sigma$. Both distributions have a peak near $\sim300$ au, then gradually fall out to thousands of au. In the analytical model, the deficit of systems with low companion separations ($\sim50-300$ au) relative to the initial conditions can be attributed to a high rate of tidal disruption at high eccentricities, as calculated in panel (c) of Figure \ref{fig:analytic_model}. While this deficit appears in the model population when not accounting for mass loss, it becomes especially strong when systems that undergo mass loss are included. The observations similarly reveal a significant dearth of close-in stellar companions. Although \cite{Thebault+25} do not exclusively focus on hot Jupiters, they find a peak in the entire exoplanet companion separation distribution at $\sim500$ au with $>3\sigma$ significance, suggesting that the peak in the distribution here may be a real feature. This distribution is significantly different from the sample of field star separations, which has a peak near 50 au \citep[e.g.,][]{Duquennoy+91,Raghavan+10,Kraus+16}. While our study avoids $<50$ au separations where planet formation is most strongly suppressed \citep[e.g.,][]{Kraus+16}, it is possible that suppression of planet formation plays a confounding role in the deficit of systems with low companion separations \citep[e.g.,][]{Moe+21,Sullivan+26}. Future work to more tightly constrain companion separation distributions in the cold and hot Jupiter populations may help disentangle the effects of initial planet formation suppression and later-stage tidal disruption.

\subsection{Period distribution}

\begin{figure*}[ht]
\begin{center}

\includegraphics[width=6.5in]{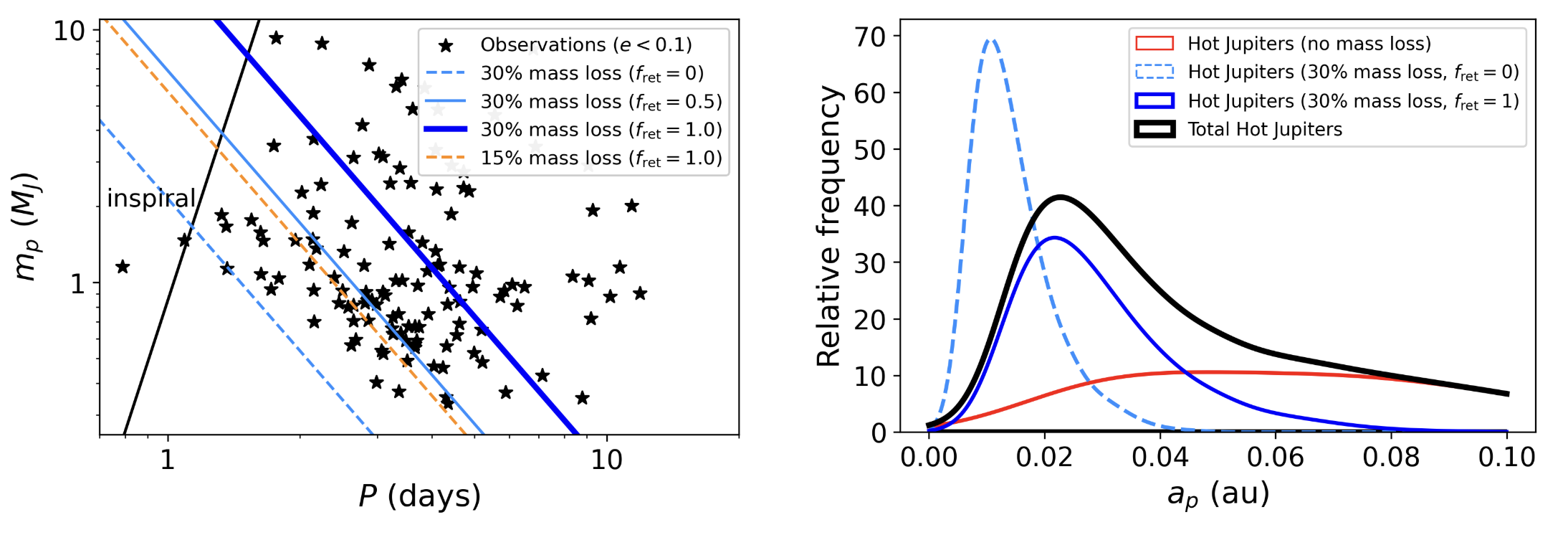}
\caption{\footnotesize Left: Mass-period distribution for observed hot Jupiters (black stars) that are relatively circularized (measured eccentricities $<0.1$). The observed sample of hot Jupiters is taken from the NASA Exoplanet Archive (accessed August 22, 2025) \citep[e.g.,][]{Akeson+13,PSCompPars}. We select systems with $0.3M_J<m_p<10M_J$ and $a_p<0.1$ au. The analytical thresholds from Eq.~(\ref{eq:final_a}) are shown for 30\% mass loss, with $f_{\rm ret} = 0$ (dashed light blue line), $f_{\rm ret} = 0.5$ (solid light blue line), and $f_{\rm ret} = 1.0$ (solid dark blue line). We also show the threshold for 15\% mass loss with $f_{\rm ret} = 1.0$. The threshold for inspiral and merger from stellar tidal dissipation is shown with a solid black line, calculated using the calibration from \cite{Hansen10}. Right: Semi-major axis distribution for hot Jupiters from the analytical model, smoothed using a Gaussian kernel. The solid red histogram shows systems that become a hot Jupiter with no mass loss. The dashed light blue line shows systems that undergo loss with $f_{\rm ret} = 0$, and the solid dark blue line shows the same systems with $f_{\rm ret} = 1$ (assuming 30\% mass loss). The total hot Jupiter population from the model (both with and without mass loss) is shown with a solid black line.}
\label{fig:massperiod_analytic}
\end{center}
\end{figure*}

It has been noted before that relaxing the disruption criterion can significantly improve the rate of hot Jupiter production \citep[e.g,][]{Naoz+12,Petrovich15b}. However, these models did not account for angular momentum redistribution following mass loss. We next analytically explore how this angular momentum return affects the final location of hot Jupiters.
 
For a planet that loses mass $\Delta m_p$, the returned angular momentum to the planetary remnant of mass $m_{p,f} = m_p - \Delta m_p$ is $f_{\rm ret} \Delta m_p h_0$, where $h_0$ is the specific angular momentum at the time of mass loss. The final angular momentum of the planet following the angular momentum return is  
\begin{equation}
    h_f m_{p,f} = h_0 (m_{p,f} + f_{\rm ret}\Delta m_p) \\= h_0 (f_{\rm ret}m_{p,0}+ (1-f_{\rm ret})m_{p,f} ).
\end{equation}
The final location of a planet following circularization along a track of nearly constant angular momentum is \citep[e.g.,][]{Ford+05}
\begin{equation}
    a_{f} = a(1-e^2)\sim 2a(1-e) = 2q \ ,
\end{equation}
where the approximation holds for $e \to 1$. Given that $a \propto h^2$, the final circularization location of the planet is shifted outward by a factor $(f_{\rm ret}m_{p,0}/m_{p,f}+ (1-f_{\rm ret}))^2$. For systems that approach near the Roche limit, this gives 
\begin{align}
a_f &\sim 9.36 \times 10^{-3} \, {\rm au}
\left(\frac{q}{r_t}\right)
\left(\frac{R_p}{R_J}\right)
\left(\frac{m_{p,0}}{M_J}\right)^{-1/3}
\left(\frac{m_*}{M_\odot}\right)^{1/3} \notag \\
&\quad \times \left(f_{\rm ret} \frac{m_{p,0}}{m_{p,f}} + (1 - f_{\rm ret})\right)^2 \ ,\label{eq:final_a}
\end{align}
which reduces to prior results for $q/r_t = 2.7$ and $f_{\rm ret}=0$. This location depends on the planet radius, final mass, and pericenter obtained, which are not strictly known \textit{a priori} as they depend on the details of the orbit-by-orbit response to mass loss. However, the formula can broadly be used to approximate the location for reasonable estimates of these quantities. 

We demonstrate this effect at the population level in Figure \ref{fig:massperiod_analytic} and compare with observations. The observed sample of hot Jupiters is taken from the NASA Exoplanet Archive (accessed August 22, 2025)\citep[e.g.,][]{Akeson+13,PSCompPars}. We select systems with $0.3M_J<m_p<10M_J$, $a_p<0.1$ au, and $e_p<0.1$. We look at planets that are relatively circularized to provide a faithful comparison to the final states in the model, as more eccentric planets may still be in the process of migrating from higher semi-major axes. To construct a clean sample, only systems with well-measured eccentricities are included (i.e., upper and lower error bars are reported on the measurement). Altogether, 127 observed planets are represented in our sample. 

In the left panel, we show the population of observed hot Jupiters. The threshold for inspiral and merger from stellar tidal dissipation is shown with a black line, calculated using the calibration from \cite{Hansen10}
\begin{equation}
\label{eq:hansen}
    a_{\rm inspiral} \sim 0.02 \, {\rm au} \, \left(\frac{m_p}{M_J}\right)^{1/8} \left(\frac{R_*}{R_\odot}\right)^{5/4} \left(\frac{T_{\rm inspiral}}{8.2 \times 10^9 \, \rm yr}\right)^{1/8}  \ ,
\end{equation}
where in the figure we take $m_p = 1 M_J$, $R_* = 1R_{\odot}$ is the stellar radius, and $T_{\rm inspiral} = 8.2 \times 10^9 \, \rm yr$ as a characteristic inspiral time. We also show the final location thresholds from Eq.~(\ref{eq:final_a}) for varying levels of mass loss and angular momentum return. These thresholds align with the bulk of the inner hot Jupiter population, suggesting that angular momentum return from mass loss may be at play in driving some of the observed planets outward. While the disruption threshold with $f_{\rm ret} =0$ naturally carves out the lower portion of the population, it does not capture the final location of the majority of planets. Here, we have shown curves of fixed loss for all planets for illustrative purposes. Obtaining more accurate values for mass loss in each individual system requires the more extensive numerical framework. 

In the right panel of Figure \ref{fig:massperiod_analytic}, we show the semi-major axis distribution for the population from the fiducial Monte Carlo analysis. The dashed light blue line shows the final locations of systems that undergo mass loss with
$f_{\rm ret} = 0$, and the solid dark blue line shows the same systems with $f_{\rm ret} = 1$, calculated with Eq.~(\ref{eq:final_a}). The sharply peaked distribution naturally broadens and shifts to higher $a_p$ when accounting for angular momentum return, a feature that will be explored in more detail in the numerical simulations. Hot Jupiters that undergo no mass loss (solid red line) represent a minority of the total hot Jupiter population (solid black line). To isolate the effect of angular momentum return, we have not removed systems due to stellar tidal dissipation, which would further sculpt the inner edge of the distribution. We show these distributions to highlight the key effects in the analytical model: stripped
planets constitute the majority of hot Jupiter systems,
and when angular momentum return is accounted for,
the peak of the final semi-major distribution shifts to higher
values.

\section{Numerical Model}
\label{sec:numerical}

We next perform detailed numerical modeling to track the time evolution of systems and study the response of the planet's structure and orbit to mass loss. These simulations enable more robust comparisons to the observed hot Jupiter population.

\subsection{Secular evolution}

We numerically solve the octupole-level secular equations for the hierarchical three-body system following \citet{Naoz+11,Naoz+13}. In addition, we include general relativistic precession of the inner and outer orbit \citep[][]{Naoz+13b}. Equilibrium tides are also included following \cite{Eggleton98} and \cite{Eggleton+01}. We fix the viscous timescale $t_{v,1} = 50$ years for the star and $t_{v,2} = 0.01$ years for the planet. These values are informed by previous population syntheses calibrated to real systems \citep[e.g.,][]{Petrovich15b}. We define the stellar obliquity $\psi$ as the angle between the spin of the inner star and the direction of the angular momentum of the inner orbit. The sky-projected obliquity $\lambda$ can be measured through the Rossiter-McLaughlin effect \citep[e.g.,][]{Gaudi+07}. Here, the stellar and planetary spins are initialized to the solar (24.47 days) and Jovian (0.41 days) values, respectively, with spin–orbit alignment ($\psi=0.001^\circ$)\footnote{We ultimately find that the mass loss prescription in this work has a negligible effect on the final stellar obliquity distribution, as discussed in Appendix \ref{app:obliquities}.}. Our code follows the precession of the spin vector \citep[again see for full set of equations,][]{Naoz16}. Furthermore, we model the stellar evolution of the stars using the SSE evolution code of \cite{Hurley2000}. The combined code with secular evolution, general relativity, tides, and stellar evolution has been tested and applied to various astrophysical systems \citep[e.g.,][]{Naoz16,Naoz+16,Stephan+16,Stephan+18,Stephan+19,Stephan+20,Stephan21,Angelo+22,Shariat+23,Shariat+24,Weldon+25,Holzknecht+25}.

The upper limit for each system's integration time in our simulations is approximately the main sequence lifetime of the host star, using $t_{\rm lifetime} \propto m_*^{-2.5}$ (maximum of 10 Gyr). We stop the simulation if a planet is engulfed during stellar evolution, or if the planet's mass falls below $0.3 M_J$ following mass loss (i.e., the planet is no longer classified as a giant). If a hot Jupiter forms, the orbit circularizes until the integration is stopped once $e_p < 10^{-4}$.

\subsubsection{Mass loss}

We explore various models for mass loss as a function of periapse distance, owing to the stochastic nature of mass loss found in the hydrodynamical simulations of \cite{Faber+05} and \cite{Guillochon+11}. We begin by using the first-passage results of \cite{Guillochon+11} (see their Figure 10) in order to determine the fractional mass loss as a function of pericenter distance during each passage. This amounts to treating the stripped planet as having no interaction with stripped debris on subsequent passages. In the same manner, \cite{Yu+24} used the results of \cite{Guillochon+11} to estimate mass loss during close encounters in the context of secularly-driven migration to explain the properties of the planet WASP-107 b. Using a similar approach, we fit an exponential to the fractional mass loss in \cite{Guillochon+11} and obtain
\begin{equation}
    \frac{\delta m_p}{m_p} = A \exp\left(-B \frac{q}{r_t}\right)  \ ,
\end{equation}
where $A = 8 \times 10^6$ and $B = 11.5$ \citep[using][]{WebPlotDigitizer}. This will be referred to as the "moderate loss" case, as we will explore both higher and lower mass loss fractions. To ensure a numerically smooth transition from the regime of no mass loss, in the code, we take 
\begin{equation}
\label{eq:massloss}
    \frac{\delta m_p}{m_p} = A \exp\left(-B \frac{q}{r_t}\right) \left[ 1+\exp \left(250 \left(\frac{q}{r_t}-2.7 \right) \right)\right]\ .  
\end{equation}
Planets may pass from $q/r_t>2.8$ (no mass loss occurs) to $2.7<q/r_t<2.8$ (a negligible, but smooth amount of loss occurs), to $q/r_t<2.7$ (the mass loss follows the exponential fit). 

Compared to \cite{Guillochon+11}, \cite{Faber+05} found that planets may approach closer ($q/r_t<2.2$) before losing mass, then the mass loss steeply rises before following a similar exponential form to \cite{Guillochon+11}. This result indicates that mass loss may proceed more gently than in \cite{Guillochon+11}. To explore the case of more gentle mass loss, we take $A = 5 \times 10^{4}$ and $B = 11.5$ in Eq.~(\ref{eq:massloss}). We refer to this as the ``low loss'' case. 

To account for a scenario in which planets undergo some interaction with stripped material, akin to the later-passage mass loss fractions of \cite{Guillochon+11}, we also consider a ``high loss'' case, in which $A = 5 \times 10^8$ and $B = 11.5$. This is a higher fraction of mass loss than explored in \cite{Yu+24}.

Mass-losing interactions during each individual periapse passage make secular orbit averaging difficult. Many orbits and episodes of mass loss may occur within a large secular time step. Therefore, when a planet falls below $q/r_t = 2.8$ and mass loss becomes possible, we set the integration time step equal to the orbital period of the planet. Taking a smaller time step enhances the resolution of the simulation to individual orbits. This approach allows for the calculation of the planetary and orbital parameters following mass loss on an orbit-by-orbit basis as mass is lost during each periapse passage. Following each episode of mass loss, we treat the specific angular momentum $h$ as conserved to recalculate the orbit of the planet, assuming that the mass is lost at periapse \citep[see e.g.,][for similar treatment of mass loss in hierarchical triple systems]{Lu+19}. Differing from the treatment of \cite{Yu+24} that fixes the radius during mass loss, we adjust the radius of the planet assuming that the central entropy is constant during a rapid mass loss event, a procedure described in more detail in Section~\ref{sec:MESA}. This approach allows for modeling of the interplay between mass loss and planetary and orbital response at the resolution of individual orbits. We are able to determine whether mass loss makes certain planets more or less susceptible to additional mass loss on subsequent orbits. 

\subsubsection{Angular momentum return}

For each parcel of lost mass $\delta{m_p}$ lost during an individual orbit, a fraction $f_{\rm ret}$ is retained by the system and ultimately accreted onto the star ($\delta{m_p} \ll m_*$). This accretion is
facilitated by torques between the tidal streamers and the planetary orbit, so that a fraction $f_{\rm ret}$ of angular momentum is returned to the planetary orbit by taking
\begin{equation}
    \frac{dh}{dt} = f_{\rm ret} \frac{\delta{m_p}}{m_p} \frac{r}{r_0} \frac{h_0}{\tau}  \ ,
\end{equation}
where we set $\tau = 100$~yr to be a characteristic timescale for the angular momentum to return\footnote{The value of $\tau$ can be motivated heuristically from the viscous timescale of the accretion flow \citep[e.g.,][]{Shakura+73}
\begin{equation}
    \tau = \frac{L^2}{\alpha c_s h} \ ,
\end{equation}
where $L$ is the characteristic length scale of the flow (we take $L \sim 20R_J$ from visualizations in \cite{Guillochon+11}), $\alpha$ is a dimensionless parameter that captures uncertainty in the mechanisms of angular momentum transport (typically $\alpha = 10^{-4} -1$, we take $\alpha \sim 0.01$), $c_s$ is the sound speed ($c_s \sim 10^6$ cm/s for $\sim 10^4$ K gas), and $h$ is the scale height (we take $h \sim 0.1 R_J$). Many of these factors are poorly understood, and $\tau$ can vary significantly depending on the time-varying geometry of the tidal streamers and the nature of the stripped gas. Therefore, this value is chosen to heuristically explore angular momentum return over secular timescales. We manually verified that the overall trends are relatively insensitive to $\tau$ from $\sim10-10^5$ yr, as expected given that these timescales are shorter than characteristic EKL timescales \citep[e.g.,][]{Antognini15}.}, $r = a_p(1+e_p^2/2)$ is the time-averaged lever arm for the tangential torque applied on an eccentric orbit, assuming the torque is applied on a secular timescale and is therefore constant over timescales longer than individual orbits. $r_0$ is the lever arm at the time of mass loss, and $h_0$ is the specific angular momentum at the time of mass loss. Over multiple orbits, the torque from many parcels of lost mass accumulates. We denote the cumulative lost mass that a planet experiences over many parcels as $\Delta m_p = \sum \delta m_p$. We stop applying a parcel's torque after the angular momentum of that parcel $f_{\rm ret} \delta m_p h_0$ is completely given back to the orbit. This work differs from previous studies that treat all angular momentum return as happening at periapse \citep[e.g.,][]{Sepinsky+07}. Here, we return the angular momentum considering the orbit-averaged torques that are exerted over secular timescales.

\subsection{Planetary Structure}

Under the conditions of mass transfer, the response of the mass-losing body can have a significant effect
on the long-term evolution of the system, as it will determine the degree to which the donor object continues
to overflow its Roche lobe. We therefore model this effect in our simulations.

\label{sec:MESA}
\subsubsection{MESA models}

To account for the planetary response to mass loss, we use Modules for Experiments in Stellar Astrophysics (MESA)
\citep[e.g.,][]{Paxton2011, Paxton2013, Paxton2015, Paxton2018, Paxton2019}, which has the ability to model the evolution of giant planets. We construct grids of giant planet models that are used to inform radius contraction due to cooling and the radius response to tidal mass loss. While, in principle, it is possible to evolve MESA and the secular code simultaneously in an iterative process \citep[e.g.,][]{Gao+25}, such simulations are computationally prohibitive for large population synthesis studies, especially when accounting for mass loss on the resolution of individual orbits. Therefore, the use of pre-calculated grids allows us to approximate the structural evolution for many systems over a large parameter space.

The planet models in this work are constructed with a 30$M_{\oplus}$ rocky core and a gaseous envelope of solar composition. Traditionally, smaller cores of $\sim10-20$$M_{\oplus}$ are taken in studies of Solar System giants \citep[see for review e.g.,][]{Guillot05}, but recent observations and modeling work have suggested extrasolar gas giants can have core masses upwards of 50$M_{\oplus}$ 
\citep[e.g.,][]{Thorngren+16,Millholland+20,Wang25,Chachan+25}; we therefore take an intermediate value. This choice is likely to have a minimal impact on our results, given the similar mass-radius relations reported by other works for lower core masses \citep[e.g.,][]{Paxton2013}. However, the full effects of varying core mass on dynamical evolution will be explored in future work. 

We vary the total planet mass, using masses of 0.3-0.9$M_J$ (spaced in increments of 0.1$M_J$) and 1.0-10.0$M_J$ (spaced in increments of 0.5$M_J$). We also vary the stellar irradiation levels, considering fluxes of $10^{4}$, $5 \times 10^{4}$, $10^{5}$, $5 \times 10^{5}$, $10^{6}$, $5 \times 10^{6}$, $10^{7}$, and $5 \times 10^{7}$ erg s$^{-1}$ cm$^{-2}$. These values span the range of fluxes experienced over the initial conditions. We additionally perform a run with 1 erg s$^{-1}$ g$^{-1}$ injected uniformly over the planet (and stellar flux of $10^7$ erg s$^{-1}$ cm$^{-2}$). The latter models with heating are not used to directly model the time evolution of a planet subject to additional heating. Rather, these models provide access to the mass-radius relation at high internal entropy for planets that may undergo mass loss at early times.

We use the central entropy $S$ of the planet at a given time, along with the planet mass, to find the planet radius \citep[e.g.,][]{Hubbard77,Arras+06,Hallatt+25b}. We construct mass-radius relations from the MESA models and show these in Appendix \ref{app:mesa_cooling}.

\subsubsection{Structural evolution}

Planets cool over time as they radiate heat from their formation, causing their radius to contract. We model this radius contraction before any potential episodes of tidal mass loss and between episodes of mass loss, as the structure of the planet will ultimately determine its susceptibility to tidal disruption. 

The rate of cooling depends on the incident stellar flux. The time-averaged flux a planet on an eccentric orbit receives is 
\begin{equation}
    F = \frac{L_*}{4 \pi a_p^2 \sqrt{1-e_p^2}}\label{eq:flux} \, .
\end{equation}
For the stellar luminosity given the initial stellar mass, we use the empirical relation $L_*/L_{\odot} = (M_*/M_{\odot})^{3.5}$ \citep[e.g.,][]{Harwit88}. A planet undergoes EKL oscillations on timescales that are characteristically shorter than the Kelvin-Helmholtz timescale to radiate thermal energy. Therefore, we treat the stellar flux as constant throughout the evolution. However, we consider that on long timescales, the time-averaged eccentricity of the planet boosts the amount of flux received by a factor of $(1-e_p^2)^{-1/2}$. As we are evolving planets that cycle between $e_p \sim 0$ and $e_p \sim 1$, we use Eq.~(\ref{eq:flux}) with an approximate average eccentricity of $e_p \sim 0.5$ along with the planet's initial semi-major axis to estimate the stellar irradiation level in the MESA grid. This choice of average eccentricity effectively amounts to boosting the flux by $\sim15$\%. This boosting factor becomes significant only for very high eccentricities, where planets spend little time, so we do not expect our results to change significantly for other reasonable choices of time-averaged eccentricities.

We fit the entropy evolution of the MESA models as they cool over time, as a function of planet mass and stellar flux (see Appendix \ref{app:mesa_cooling} for procedure). Given the fitted entropy $S(t)$, we can use the MESA mass-radius relations to find the radius of the planet over time. A planet that loses mass almost
instantaneously will adjust its radius on a dynamical timescale \citep[e.g.,][]{Paczynski+72}. This timescale is far too
short for significant cooling to occur, and giant planets are almost fully convective and hence isentropic throughout \citep[e.g.,][]{Arras+06}.
As a planet loses mass, it will adjust its radius to that of a planet with the changed mass but the same internal entropy. Following an episode of mass loss, the planet begins cooling again along a new evolutionary track. Stripped planets, therefore, have different radii than undisturbed planets and appear to have an effective age that is different from their actual age. To obtain this effective age, we invert the fitting procedure by providing the new mass, stellar flux, and entropy to find the time along the new cooling curve with different fit parameters.

As the planet's orbit eventually shrinks during tidal migration, the stellar flux on the planet increases. This work is predominantly focused on the survival of planets through highly eccentric passages at nearly constant semi-major axis and thus nearly constant time-averaged flux. Therefore, the treatment is appropriate for the phase of evolution studied in this work. However, after this phase, the increased stellar flux likely plays an insignificant role on the radius evolution. Previous studies show that surface heating mostly goes into increasing the size of the isothermal surface layer, rather than raising the interior adiabat \citep[e.g.,][]{Wu+13,Komacek+17,Komacek+20}. Indeed, these studies have been largely unable to explain the anomalously large radii of some hot Jupiters with increased surface heating alone \citep[for a review, see][]{Thorngren+24}. Therefore, our approximation of constant stellar irradiation is unlikely to have a significant effect on the system outcomes.

There has been much recent interest in modeling planetary structural evolution in tandem with orbital evolution, as the interplay between these effects can greatly influence system architectures. In particular, radius inflation due to tidal heating has been coupled with high-eccentricity migration to potentially explain individual systems of interest and features in the giant planet population \citep[e.g.,][]{Rozner+22,Glanz+22,Yu+24,Lu+25,Gao+25,Hallatt+25b}. Though it remains uncertain where tidal heating is deposited in giant planets, such heating may occur in the deep interior of planets and significantly affect the system evolution. However, tidal heating generally becomes significant only after a planet has survived initial eccentric passages, during the ``warm'' tidal capture phase in which the semi-major axis is reduced by tidal dissipation. The response of the planet to mass loss during the initial "cold" passages of high eccentricity---a process that destroys a substantial fraction of source planets---remains to be addressed in detail. Tidal heating may lead to late episodes of mass loss and remove some fraction of planets from the population, a possibility that will be explored in a future study. It is first necessary to understand the survival of planets through the highly eccentric passage phase, which is the focus of this work.

\begin{figure*}[ht]
\begin{center}

\includegraphics[width=6.5in]{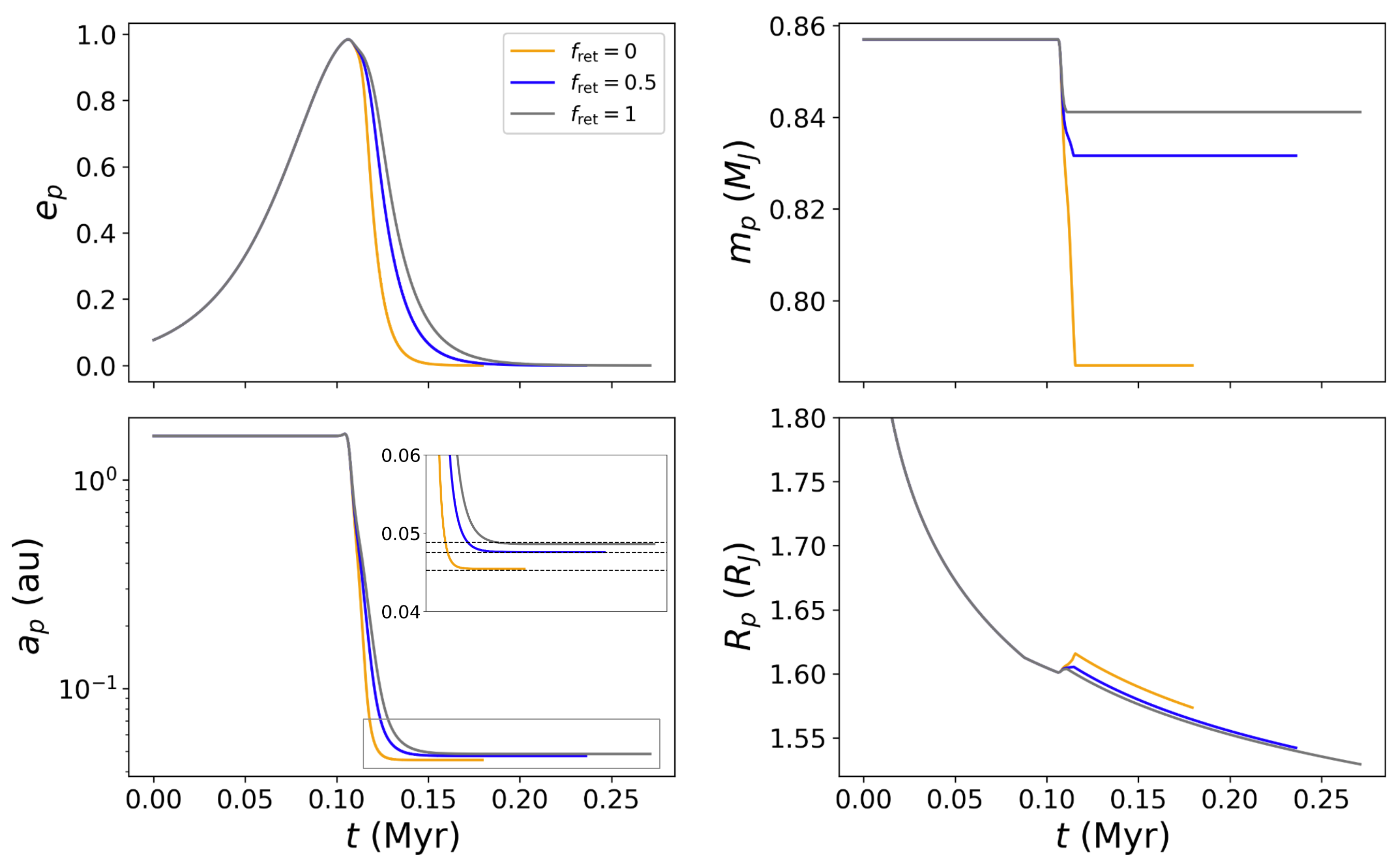}
\caption{\footnotesize Time evolutions of $e_p$ (top left), $a_p$ (bottom left), $m_p$ (top right), and $R_p$ (bottom right) for an example system that forms a hot Jupiter through one quadrupole cycle. Evolutions with varying fractions of angular momentum return are shown, including $f_{\rm ret} = 0$ (orange curves), $f_{\rm ret} = 0.5$ (blue curves), and $f_{\rm ret} = 1$ (gray curves). The $a_p$ panel (bottom left) shows a zoomed-in view of the final circularization locations of the hot Jupiter (inset panel corresponds to the gray rectangle), with the dashed black lines corresponding to analytical predictions from Eq.~(\ref{eq:final_a}). This system initially has $m_* = 1.427 M_{\odot}$, $m_p = 0.857 M_J$, $m_c = 0.838 M_{\odot}$, $a_p = 1.621$ au, $a_c = 82.846$ au, $e_p = 0.077$, $e_c = 0.084$, $\omega_1 = 46.075^{\circ}$, $\omega_2 = 212.893^{\circ}$, and $i = 84.210^{\circ}$.} 
\label{fig:example_t}
\end{center}
\end{figure*}

\subsection{Simulation runs}

We draw 2000 systems from the initial conditions described in \ref{sec:initialconditions}. As a baseline, we evolve the 2000 systems using the previous treatment that includes no mass loss (labeled \texttt{NoLoss}). In this run, 5.8\% become hot Jupiters and 16.6\% are considered tidally disrupted, consistent with \cite{Petrovich15b} for similar initial conditions. Our goal is to refine these occurrence rates using improved physical models. Most of the 2000 systems are stable against EKL oscillations and never undergo significant eccentricity evolution. These systems are therefore not affected by the new treatment for highly eccentric passages. Since re-running these would add substantial computational cost without yielding new insight, we restrict our attention to systems that do reach high eccentricities, where differences between the old and new prescriptions are expected.

Specifically, we focus on systems that in the baseline run reach a pericenter $<10r_t$. In the original setup, where planets had a fixed radius of $1R_J$, a threshold of $5r_t$ included all systems that either became hot Jupiters or suffered tidal disruption, along with some that reached high eccentricities without migrating. In the new treatment, the radius of the planet is generally $< 2 R_J$, so a threshold of $10 r_t$ safely encompasses additional systems that may undergo strong tidal interactions. A total of 573 systems fall within the $10 r_t$ threshold. Although the simulation runs in this work only evolve these 573 systems, we calculate occurrence rates considering the entire initial population of 2000 planets. For this set of systems, we explore nine different possibilities: low loss, moderate loss, and high loss (each with $f_{\rm ret} = 0$, $f_{\rm ret} = 0.5$, and $f_{\rm ret} =1$).

The initial conditions of the aforementioned runs are chosen to provide a direct comparison to prior models \citep[e.g.,][]{Weldon+25}; we call this set of initial conditions "Set 1." To explore the sensitivity of the results to the choice of initial conditions and compute more refined occurrence rates, we perform an additional run with a second set ("Set 2") of 2000 systems. In the simulation run \texttt{M0.5set2}, instead of drawing the stellar masses $m_1$ and $m_3$ independently, we use Figure 9 of \cite{ElBadry+19} to correlate the masses \citep[using][]{WebPlotDigitizer}. As a result, planet-hosting stars are more likely to have a companion similar in mass. Additionally, while we restrict the host mass $0.6 M_{\odot}<m_1<1.6M_{\odot}$ (roughly FGK stars), companions may fall outside of this range. Therefore, we allow $0.1 M_{\odot}<m_3<2.5M_{\odot}$. Finally, although the present-day cold Jupiter semi-major axis distribution is not necessarily indicative of the initial distribution, we take the observed distribution of cold Jupiters from 2-8 au from \cite{Fulton+21}. We sample $a_1$ from their parametric distribution 
\begin{equation}
    p(a_1) \propto 350 \left(\frac{a_1}{\rm au}\right)^{-0.86} (1-e^{-(a_1/3.6 \, \rm{au})^{1.59}}) \, .
\end{equation}
The \texttt{M0.5set2} run takes the fiducial choices of the moderate loss prescription and $f_{\rm ret}=0.5$.

\subsection{Example system}

To illustrate how the approximations affect system evolution, we show an example system that undergoes tidal mass loss and forms a hot Jupiter in Figure \ref{fig:example_t}. The upper left panel shows $e_p$, the upper right panel shows $m_p$, the lower left shows $a_p$, and the lower right shows $R_p$. The system is evolved under the "moderate loss" prescription, and each curve corresponds to a varying fraction of angular momentum return. We consider $f_{\rm ret}=0$ (orange curves), $f_{\rm ret}=0.5$ (blue curves), and $f_{\rm ret}=1$ (gray curves). 

For this system, $e_p$ increases to near unity due to the EKL perturbations from the companion star. At high eccentricity, the planet begins losing mass, and the orbit-by-orbit response of the planet to mass loss is calculated. The planet radius contracts prior to mass loss following the MESA grid cooling curve, adjusts adiabatically during mass loss, and begins cooling along a new cooling curve after mass loss. At high eccentricity, tidal dissipation becomes sufficiently strong to drag the planet down onto a close-in, circularized orbit. This system represents the simplest case, in which the drag-down occurs over a single quadrupole cycle. We select this system for illustrative purposes. Some systems in the population studies undergo multiple quadrupole cycles in which mass is lost before forming a hot Jupiter or undergoing tidal disruption.

We also see that varying the fraction of angular momentum return affects the final mass of the planet and its final circularization location. A higher fraction of angular momentum return leads to a lower level of planetary stripping. Following the first mass-losing orbit, the planet is pushed to higher $q$, leading the planet to experience a lower level of mass loss on the next orbit. A higher fraction of angular momentum return also pushes the final circularization location of the planet outward. Indeed, we see the $a_p$ track for $f_{\rm ret} = 1$ is higher than for $f_{\rm ret} = 0.5$, which in turn is higher than the $f_{\rm ret} = 0$ track. 

The $a_p$ panel shows a zoom-in view of the final circularization locations, with the dashed line corresponding to analytical predictions discussed in Section~\ref{sec:analytical}. To illustrate the agreement of the approximation for a simulation that has been performed, we show the analytical predictions of Eq.~(\ref{eq:final_a}) in the inset panel in the bottom left of Figure \ref{fig:example_t}. For simplicity, here we take $R_p = 1.6R_J$. Informed by the simulations, we take $q = 2.55r_t$ ($f_{\rm ret} = 0)$, $q = 2.6r_t$ ($f_{\rm ret} = 0.5)$, and $q = 2.65r_t$ ($f_{\rm ret} = 1)$, as stronger torques push the maximum attainable pericenter outward. We also take from the simulations $m_{p,f} = 0.789 M_J$ ($f_{\rm ret} = 0)$, $m_{p,f} = 0.832M_J$ ($f_{\rm ret} = 0.5)$, and $m_{p,f} = 0.841M_J$ ($f_{\rm ret} = 1)$. There is agreement between the predictions of Eq.~(\ref{eq:final_a}) and the simulations for these values. Though we have normalized this function based on the simulations, this agreement demonstrates the effectiveness of the factor in Eq.~(\ref{eq:final_a}) in capturing the general behavior of $a_f$ moving outward with the return of angular momentum.

\subsection{Occurrence rates}

\begin{table*}
\begin{center}
\setlength{\tabcolsep}{7pt}
\caption{Summary of Simulations}
\label{tab:sims}
\begin{tabular}{c c c c c c c c}
\hline\hline
Name & Initial conditions & Mass loss & $f_{\rm ret}$  & HJ (\%)$^{a}$ & TD (\%) & $f_{\rm HJ,FGK}$ (\%)$^{b}$\\
\hline
NoLoss & Set 1 & None (planets lost at $q/r_t = 2.7$) & 0 & 5.8 & 16.6 & $0.19 \pm 0.04$\\
M0 & Set 1 & Moderate - motivated by \cite{Guillochon+11}  & 0 & 12.1 & 9.2 & $0.39 \pm 0.07$\\
M0.5 & Set 1 &Moderate - motivated by \cite{Guillochon+11} & 0.5 & 15.5 & 5.7 & $0.50 \pm 0.09$\\
M1 & Set 1 &Moderate - motivated by \cite{Guillochon+11} & 1 & 16.6 & 4.6 & $0.54 \pm 0.09$ \\
L0 & Set 1 &Low & 0 & 14.8 & 6.3 & $0.48 \pm 0.08$\\
L0.5 & Set 1 &Low & 0.5 & 16.1 & 5.1 & $0.52 \pm 0.09$\\
L1 & Set 1 &Low & 1 & 17.2 & 3.8 & $0.56 \pm 0.10$\\
H0 & Set 1 &High & 0 & 9.6 & 11.7 & $0.31 \pm 0.06$\\
H0.5 & Set 1 &High & 0.5 & 12.9 & 8.4 & $0.42 \pm 0.07$\\
H1 & Set 1 &High & 1 & 14.3 & 7.0 & $0.46 \pm 0.08$\\
M0.5set2 & Set 2 &Moderate - motivated by \cite{Guillochon+11} & 0.5 & 12.2 & 6.2 & $0.39 \pm 0.07$\\

\hline \hline
\end{tabular}
\end{center}
$^{a}$The hot Jupiter (HJ) and tidal disruption (TD) fractions from the simulated population of 2000 planets.

$^{b}$The predicted occurrence rate of hot Jupiters around FGK stars calculated with Eq.~(\ref{eq:fhj}).
\end{table*}

\begin{figure*}[ht]
\begin{center}

\includegraphics[width=5in]{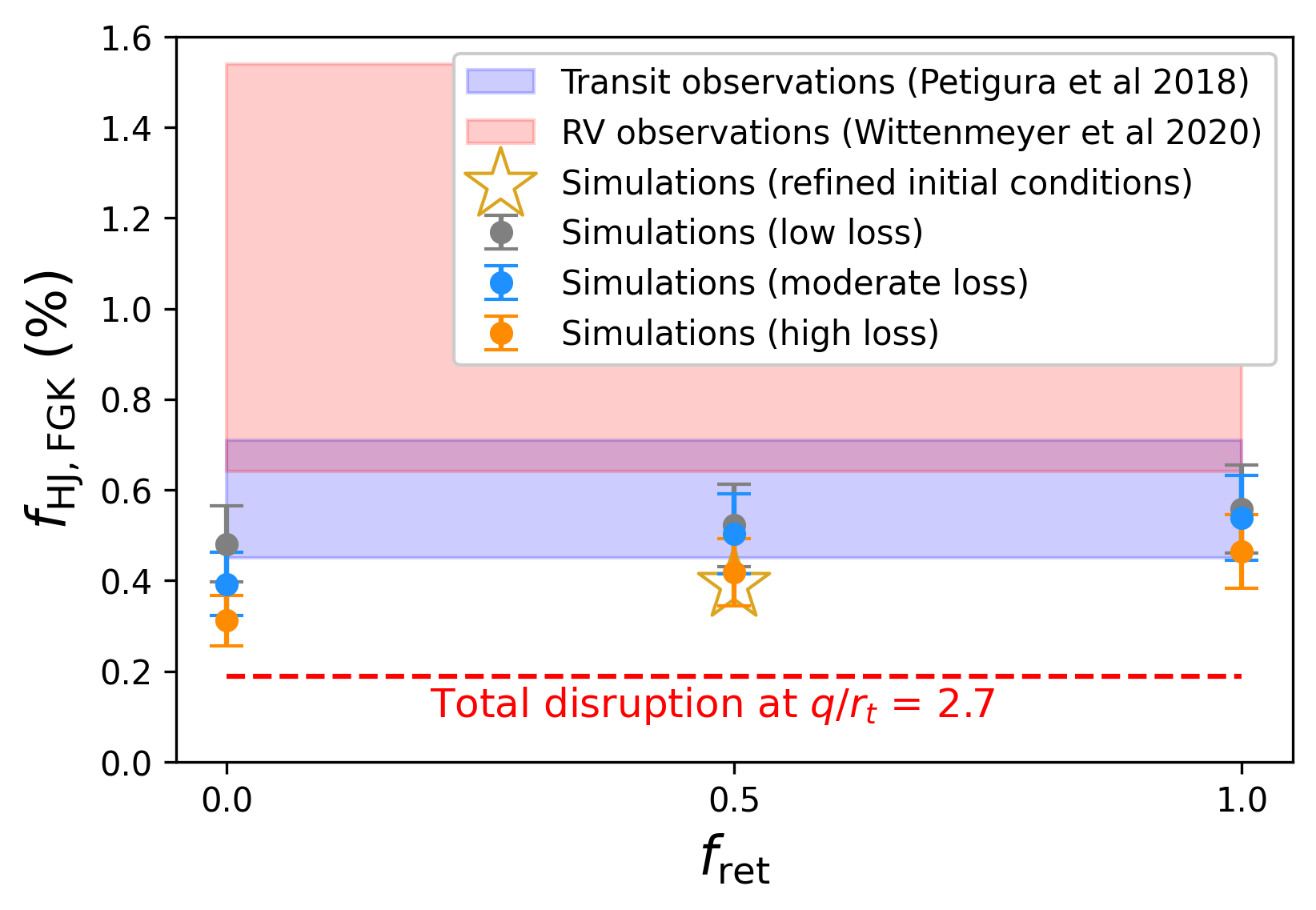}
\caption{\footnotesize Comparison of the hot Jupiter occurrence rate from the transit observations of \cite{Petigura+18} (shaded blue region), radial velocity observations of \cite{Wittenmeyer+20} (shaded red region), and those estimated from the simulations in this work. To obtain the simulated occurrence rates, we use the fraction of FGK stars in wide binaries from \cite{Raghavan+10} and the fraction of systems that harbor cold Jupiters as a source population in \cite{Fulton+21}, then multiply by the fraction of cold Jupiters that become hot Jupiters in our simulations (given in Table  \ref{tab:sims}). We show the rates for the varying mass loss prescriptions of low loss (gray points), moderate loss (blue points), and high loss (orange points), as a function of varied angular momentum return $f_{\rm ret}$. The rate for the \texttt{M0.5set2} simulations (moderate loss, $f_{\rm ret}=0.5$, "Set 2" initial conditions) is shown with a gold star. The occurrence rate estimate from the \texttt{NoLoss} run is also plotted (dashed red line), which uses the approach taken in previous studies  of removing planets that graze $q/r_t = 2.7$ \citep[e.g.,][]{Petrovich15b,Weldon+25}.}
\label{fig:rates}
\end{center}
\end{figure*}

We now turn to the population studies and summarize the results of the simulations in Table \ref{tab:sims}. For each of the mass loss procedures (moderate, low, high) and each fraction of angular momentum return ($f_{\rm ret} =$ 0, 0.5, 1) studied, we record the fraction of systems that form a hot Jupiter ($a_{1,f} < 0.1$ au) or undergo tidal disruption ($m_{p,f} < 0.3M_J$). Note that the term tidal disruption refers to systems that undergo such significant mass loss that the planet is no longer classified as a giant. This definition enables a comparison to observations with a lower mass bound of 0.3$M_J$. The potential for the planet to survive as a sub-Jovian gaseous planet or stripped rocky planet will be explored in future work.

Table \ref{tab:sims} shows the hot Jupiter formation rates and tidal disruption rates from the initial population of 2000 cold planets. It is noteworthy that, for all runs, there is a factor of $\sim2-3$ higher production rate of hot Jupiters than in the previous treatment, where planets are entirely lost at $q/r_t = 2.7$. These estimates are therefore a few times higher than in previous studies \citep[e.g.,][]{Naoz+12,Petrovich15b,Anderson+16,Weldon+25}. This boost in hot Jupiter production is comparable to the results obtained in the analytical model.

For comparison with observations, we calculate the hot Jupiter occurrence rate around FGK stars from the simulations as
\begin{equation}
    f_{\rm HJ,FGK} = {f_{\rm wide \, binary}}  \times f_{\rm CJ} \times f_{\rm HJ,EKL} \, , \label{eq:fhj}
\end{equation}
where $f_{\rm wide \, binary}$ is the fraction of FGK stars in wide binaries (or higher multiples), $f_{\rm CJ}$ is the fraction of stars harboring a cold Jupiter, and $f_{\rm HJ,EKL}$ is the simulated production rate of hot Jupiters (see Table \ref{tab:sims}). The fraction of systems in wide binaries is taken to be $23\% \pm 2\%$ \citep[][]{Raghavan+10}, considering separations greater than 50 au. This same threshold is used in our simulations, inward of which planet formation is thought to be suppressed \citep[e.g.,][]{Kraus+16}. We take $f_{\rm CJ} = 14.1\% \pm 2\%$ from \cite{Fulton+21}. We propagate the uncertainties on each measurement, along with the Poisson error in the simulations, to obtain error bars on the estimated occurrence rate from the simulations. The values of $f_{\rm HJ,EKL}$ are reported in Table \ref{tab:sims}.

Historically, transit and radial velocity surveys have reported different occurrence rates for hot Jupiters.
Radial velocity surveys have found rates of $\sim 1\%$ \citep[e.g.,][]{Mayor+11,Wright+12,Wittenmeyer+20}, whereas transit surveys estimate the rate to be a factor of $\sim2$ lower at $\sim0.5\%$ \citep[e.g.,][]{Howard+12,Fressin+13,Petigura+18}. Recent work by \cite{Beleznay+22} showed that when accounting for stellar multiplicity, the occurrence rate in the TESS sample is $0.98 \pm 0.36\%$ for G-stars, bringing transit estimates in agreement with the radial velocity estimates. For a recent discussion of hot Jupiter occurrence rates, see \cite{Gan+23}. 

Figure \ref{fig:rates} shows the observed rate and the simulated rates corresponding to different mass loss prescriptions: low loss (gray points), moderate loss (blue points), and high loss (orange points), as a function of varied angular momentum return $f_{\rm ret}$. The rate for the \texttt{M0.5set2} simulations (moderate loss, $f_{\rm ret}=0.5$, "Set 2" initial conditions) is shown with a gold star. For comparison with the observations, we show the $0.57^{+0.14}_{-0.12}\%$ estimate from \cite{Petigura+18} in the shaded blue region and the $0.84_{-0.20}^{+0.70}$ estimate from \cite{Wittenmeyer+20} in the shaded red region. These surveys are selected due to their similar stellar host samples to the simulations (FGK stars). We compare to the previous approach \citep[e.g.,][]{Petrovich15b,Weldon+25}, in which all planets are destroyed at $q/r_t = 2.7$ (dashed red line). In this previous approach, the EKL effect can only account for a small fraction of observed hot Jupiters.

The simulations in which mass loss and angular momentum return are accounted for show closer agreement with the observed rates. For the fiducial case (moderate loss, $f_{\rm ret}=0.5$), the results agree with the transit based rate and fall slightly short of the radial velocity based rate. Higher levels of angular momentum return and lower levels of mass loss enhance the production rate and produce stronger agreement. Altogether, these results suggest that stellar EKL may account for a large fraction of observed hot Jupiters, and that many observed hot Jupiters may have experienced a non-negligible degree of mass loss.

Figure \ref{fig:rates} shows that higher mass loss leads to fewer surviving hot Jupiters. This result is expected, given that tidal disruptions become more likely with a higher degree of mass loss. We also see that as $f_{\rm ret}$ increases, more hot Jupiters are able to survive due to the increased angular momentum return pushing the planets to higher $q$, where mass loss is weaker. Furthermore, the difference in production rate between the $f_{\rm ret}=0$ and $f_{\rm ret}=0.5$ cases is greater than between the $f_{\rm ret}=0.5$ and $f_{\rm ret}=1$ cases (i.e., the rate somewhat "levels off" as $f_{\rm ret}$ increases). This feature can be explained by the fact that we are accounting for mass loss and angular momentum return on an orbit-by-orbit basis, without assuming a fixed amount of mass loss. A planet with $f_{\rm ret}=1$ has a high fraction of angular momentum returned following the first episode of mass loss, pushing $q$ outwards and leading to potentially lower mass loss on subsequent orbits. On the other hand, a planet with $f_{\rm ret}=0.5$ experiences a weaker torque after the first mass-losing orbit, potentially leading to more mass loss and somewhat higher angular momentum return on subsequent orbits. These effects compete over many episodes of mass loss, though the cumulative angular momentum return effect for $f_{\rm ret}=1$ remains stronger than for $f_{\rm ret}=0.5$. This effect can also be seen in Figure \ref{fig:example_t}, where the $f_{\rm ret}=0.5$ and $f_{\rm ret}=1$ cases experience similar degrees of mass loss in comparison to the $f_{\rm ret}=0$ case (though we note there are many non-linear effects at play).

Importantly, there is little difference between the results with initial condition Set 1 and initial condition Set 2. For the fiducial choices of moderate mass loss and $f_{\rm ret}=0.5$, Set 1 yields an occurrence rate of $0.50\% \pm 0.09\%$, and Set 2 yields a rate of $0.39\% \pm 0.07\%$. While the wider average planetary separations contribute to stronger secular effects in Set 2, the presence of lower mass stars in Set 2 produces slightly weaker effects overall. However, these two simulation runs produce occurrence rates that are within $1\sigma$ agreement. We therefore conclude that the results in this work are not particularly sensitive to reasonable variations in the estimates of the initial conditions (e.g., planets are initialized approximately near the snow line and empirical stellar mass distributions are used for the perturber).

\subsection{Mass loss fractions}
\label{sec:ml_frac}

\begin{figure*}[ht]
\begin{center}

\includegraphics[width=7.0in]{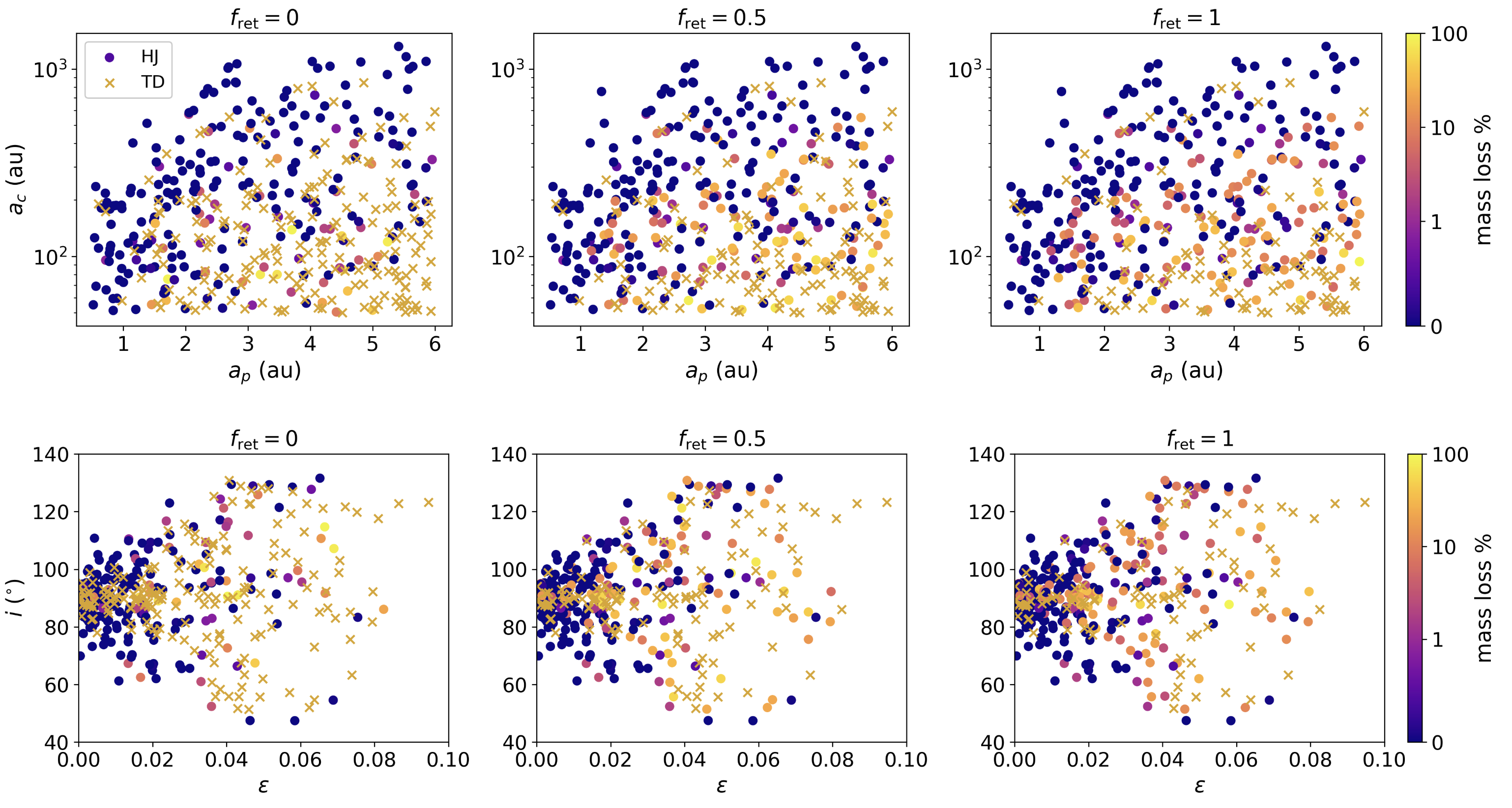}
\caption{\footnotesize Top row: Mass loss percentages as a function of initial $a_p$ and $a_c$ in the moderate loss simulations for $f_{\rm ret} =0$ (left), $f_{\rm ret} =0.5$ (middle), and $f_{\rm ret} =1$ (right). Surviving hot Jupiters are shown with circles and tidal disruptions with an orange "X". The color code corresponds to the fraction of mass lost, with warmer colors indicating higher fractions of loss. Blue circles have either no or very little mass loss, violet circles experience $\sim1$\% mass loss, orange circles experience $\sim10$\% mass loss, and yellow circles lose almost all of their mass but survive as a hot Jupiter. Bottom row: Mass loss fractions as a function of initial $\epsilon$ and $i$. The color code is the same as in the top row.}
\label{fig:masslossfractions}
\end{center}
\end{figure*}

\begin{figure}[ht]
\begin{center}

\includegraphics[width=3.5in]{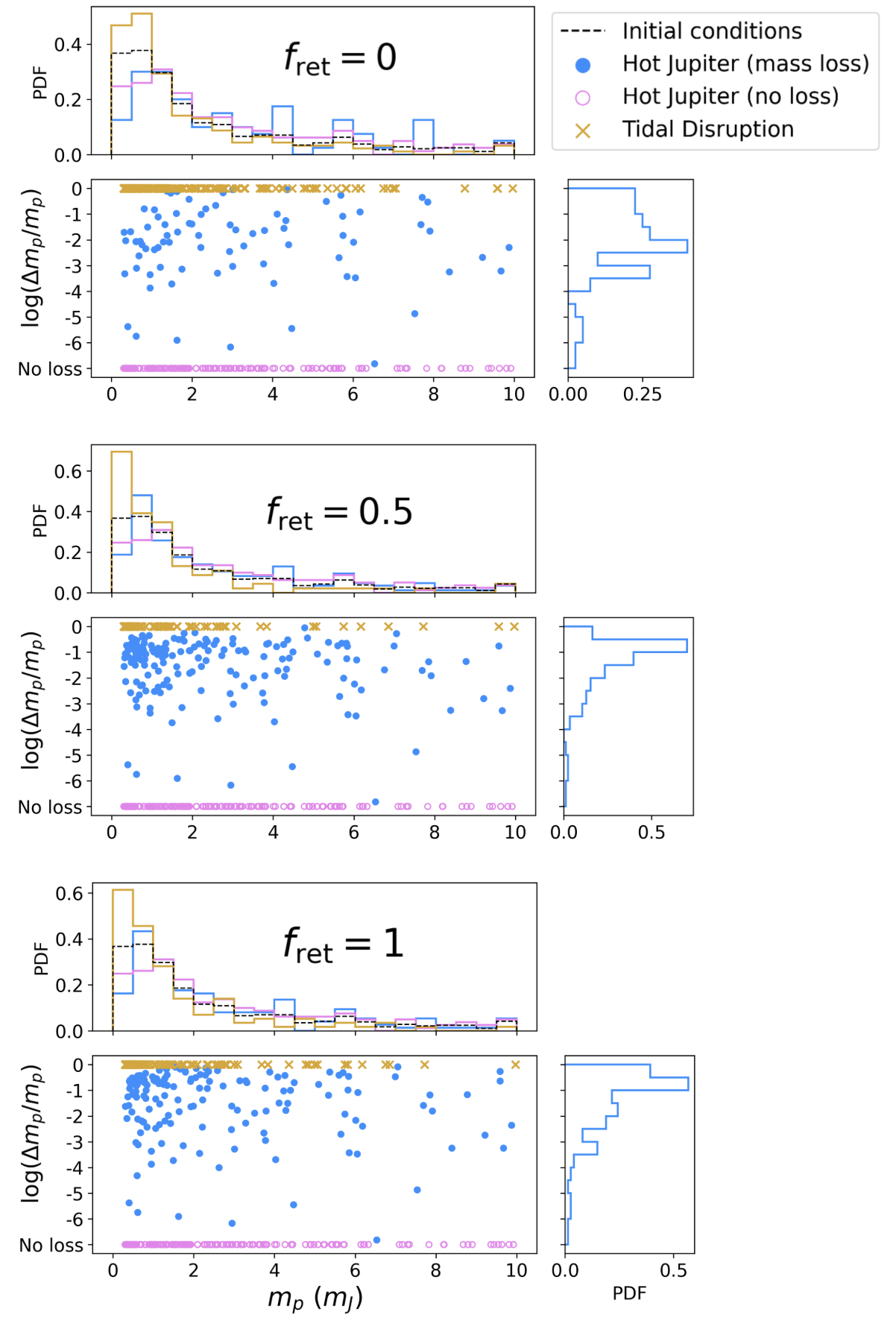}
\caption{\footnotesize Mass loss fractions as a function of initial planetary mass in the moderate loss simulations. Hot Jupiters that lose mass are shown with blue dots, hot Jupiters that undergo no mass loss with violet open circles, and tidal disruptions with an orange "X". The angular momentum return fraction is varied from $f_{\rm ret} =0$ (top), $f_{\rm ret} =0.5$ (middle), to $f_{\rm ret} =1$ (bottom). For each choice of $f_{\rm ret}$, we show the marginal histograms of the mass distributions with the same color code (top) and mass loss fractions of the hot Jupiters that lose mass (right). The initial distribution of masses is shown with a dashed black curve. \textit{We see that low mass planets undergo frequent tidal disruption. As angular momentum return "saves" planets, they survive as highly stripped hot Jupiters.}}
\label{fig:masslossfractions_mp}
\end{center}
\end{figure}

Here, we explore the degree of mass loss that planets undergo as a function of initial system conditions. The mass loss fraction is not provided to the simulations \textit{a priori}, but instead is an outcome of the complex interplay between the strength of EKL perturbations, the planet's structure and response to mass loss during each orbit, the orbit's response to mass loss, and the Roche limit's response to the changing planetary mass and radius.

The top row of Figure \ref{fig:masslossfractions} shows the mass loss fractions as a function of $a_c$ and $a_p$. Circles represent surviving hot Jupiters, and X's represent systems that lost enough mass to be considered tidally disrupted. The color bar shows the mass loss fraction, where warmer colors correspond to higher levels of mass loss. Blue circles have either no or very little mass loss, violet circles experience $\sim1$\% mass loss, orange circles experience $\sim10$\% mass loss, and yellow circles lose almost all of their mass but survive as a hot Jupiter. From left to right, the angular momentum return fraction is increased, and we see an increase in the number of surviving hot Jupiters. We only show the results for the moderate loss run to avoid clutter; however, we verified that the same general trends apply to the lower and higher loss runs. 

Planets with higher initial semi-major axes tend to experience higher mass loss, along with those that have closer-in stellar companions. This result aligns with expectations that for lower ratios of the planetary to companion semi-major axis, stronger perturbations drive the planets to plunge more deeply towards the host star and consequently lose more mass. Planets initially below $\sim1$ au rarely experience sufficiently strong perturbations to undergo mass loss. Planets initially above $\sim5$ au tend to experience high levels of mass loss. While mass loss is possible with companions out to $\sim1000$ au, systems with companions closer than $\sim500$ au tend to undergo the highest levels of loss.

In the bottom row of Figure \ref{fig:masslossfractions}, we examine the mass loss dependence on the initial mutual inclination $i$ and initial $\epsilon$, see Eq.~(\ref{eq:eps}), which describes the strength of the octupole-level perturbations \citep[see for review, e.g.,][]{Naoz16}. $\epsilon$ depends not only on the semi-major axis ratio, as studied above, but also on the eccentricity of the outer orbit. We see that for low $\epsilon$ (quadrupole-dominated), only inclinations between $\sim60-120^{\circ}$ reach sufficiently high eccentricities to form hot Jupiters.  As the octupole contribution becomes important, high eccentricity excursions are enabled between $\sim40-140^{\circ}$ \citep[as previously observed in e.g.,][]{Naoz+12}. Tidal disruptions and high amounts of mass loss occur narrowly between $\sim80-100^{\circ}$ for low $\epsilon$, and represent the majority of the outcomes for high $\epsilon$.

In Figure \ref{fig:masslossfractions_mp}, we show the mass loss fraction as a function of initial planetary mass for the moderate loss simulations with varied $f_{\rm ret}$. We verify that similar trends persist in the high and low loss simulations. Hot Jupiters that lose mass are shown with blue dots, hot Jupiters that undergo no mass loss are shown with violet open circles, and tidal disruptions are shown with an orange "X". The marginal histograms for each plot show the mass distributions with the same color code and mass loss fractions of the hot Jupiters that lose mass.

Although the distributions in part reflect the initial mass function of the planets, the mass loss fractions do exhibit a noteworthy dependence on initial planetary mass. High mass planets ($\gtrsim 5 M_J$) undergo less average fractional mass loss and less frequent tidal disruption than lower mass planets. Two physical effects are at play here. First, given the relatively flat mass-radius relation for giant planets \citep[e.g.,][]{Chen+17,Bashi+17,Muller+24}, higher mass planets are generally able to plunge more deeply before losing mass (i.e., the Roche limit radius is smaller). Second, high mass planets are more stable to mass loss. The isentropic curves for high mass planets are flatter than for low mass planets at young ages (see the MESA models in Figure \ref{fig:mesa_grids} in Appendix \ref{app:mesa_cooling}). As they lose mass, lower mass planets are more likely to increase in radius due to the steep isentropic curves, making them more susceptible to runaway mass loss on subsequent passages. Indeed, we see a high concentration of low mass planets ($\lesssim 1 M_J$) considered tidally disrupted. We note that this result is also due in part to our definition for tidal disruption (e.g., an initially $0.5M_J$ planet that loses 50\% of its mass is considered tidally disrupted by virtue of falling below 0.3$M_J$, whereas a $1M_J$ planet that loses 50\% of its mass is not). Nevertheless, we expect to see low mass planets undergo more frequent tidal disruptions and high mass planets undergo less catastrophic mass loss. 

As the angular momentum return fraction is varied in Figure \ref{fig:masslossfractions_mp}, the mass distribution of surviving planets changes. From $f_{\rm ret}=0$ to $f_{\rm ret}=0.5$, we see that many tidal disruptions of planets in the range $0.5M_J<m_p<1M_J$ become saved from tidal disruption. The angular momentum return is insufficient to save a large fraction of planets with masses below $0.5M_J$, which undergo the strongest runaway loss, but it can save many planets that are slightly more stable to loss. Turning to the marginal histograms of the mass loss fraction, we see that increasing $f_{\rm ret}=0$ to $f_{\rm ret}=0.5$ produces a large number of planets with mass loss fractions $-1<\log(\Delta m_p/m_p) < -0.5$ (i.e. $\sim$10\%-30\% mass loss). That is, many planets that get saved from tidal disruption become highly stripped hot Jupiters. As $f_{\rm ret}=0.5$ is increased to $f_{\rm ret}=1$, a broader range of $\log(\Delta m_p/m_p)$ is allowed, as the increased angular momentum return allows highly stripped planets to become weakly stripped planets.

A testable outcome of this model is that surviving hot Jupiters are, on average, different in mass than cold Jupiters. While the average mass of hot Jupiters in the simulations is reduced in mass by 2.5\% (low loss, $f_{\rm ret} = 0.5$), 6.0\% (moderate loss, $f_{\rm ret} = 0.5$), and 7.5\% (high loss, $f_{\rm ret} = 0.5$), observationally, the lowest mass planets are preferentially pushed out of the hot Jupiter population (with final masses $m_p<0.3M_J$) due to undergoing the most extreme mass loss. As a consequence, the surviving population of hot Jupiters with masses $>0.3M_J$ trends towards higher mass planets that are more likely to retain mass. At the population level, the mass of hot Jupiters relative to cold Jupiters is increased by 6.4\% (low loss, $f_{\rm ret} = 0.5$), 7.5\% (moderate loss, $f_{\rm ret} = 0.5$), and 12.6\% (high loss, $f_{\rm ret} = 1$). As demonstrated in Figure \ref{fig:masslossfractions_mp}, these effects are especially prominent in the population of planets with mass $0.3<M_J<1M_J$, so observational studies that focus on this range may place the tightest constraints on these models.

There has historically been a discrepancy between the masses of hot Jupiters and cold Jupiters, where cold Jupiters have a higher average mass than hot Jupiters \citep[e.g.,][]{Udry+03,Knutson+14,Bryan+16}. However, when accounting for selection biases that favor the detection of massive planets at large orbital periods, some surveys report that this effect is not statistically significant \citep[e.g.,][]{Zink+23}. Future observational constraints may help to resolve this tension and provide a clearer test of the models in this work. Later episodes of mass loss, induced by tidal heating or tidal inspiral, may also play a role in removing mass from the hot Jupiter population, though such studies are outside the scope of this work.

\subsection{Period distribution}

\begin{figure*}[ht]
\begin{center}

\includegraphics[width=5.4in]{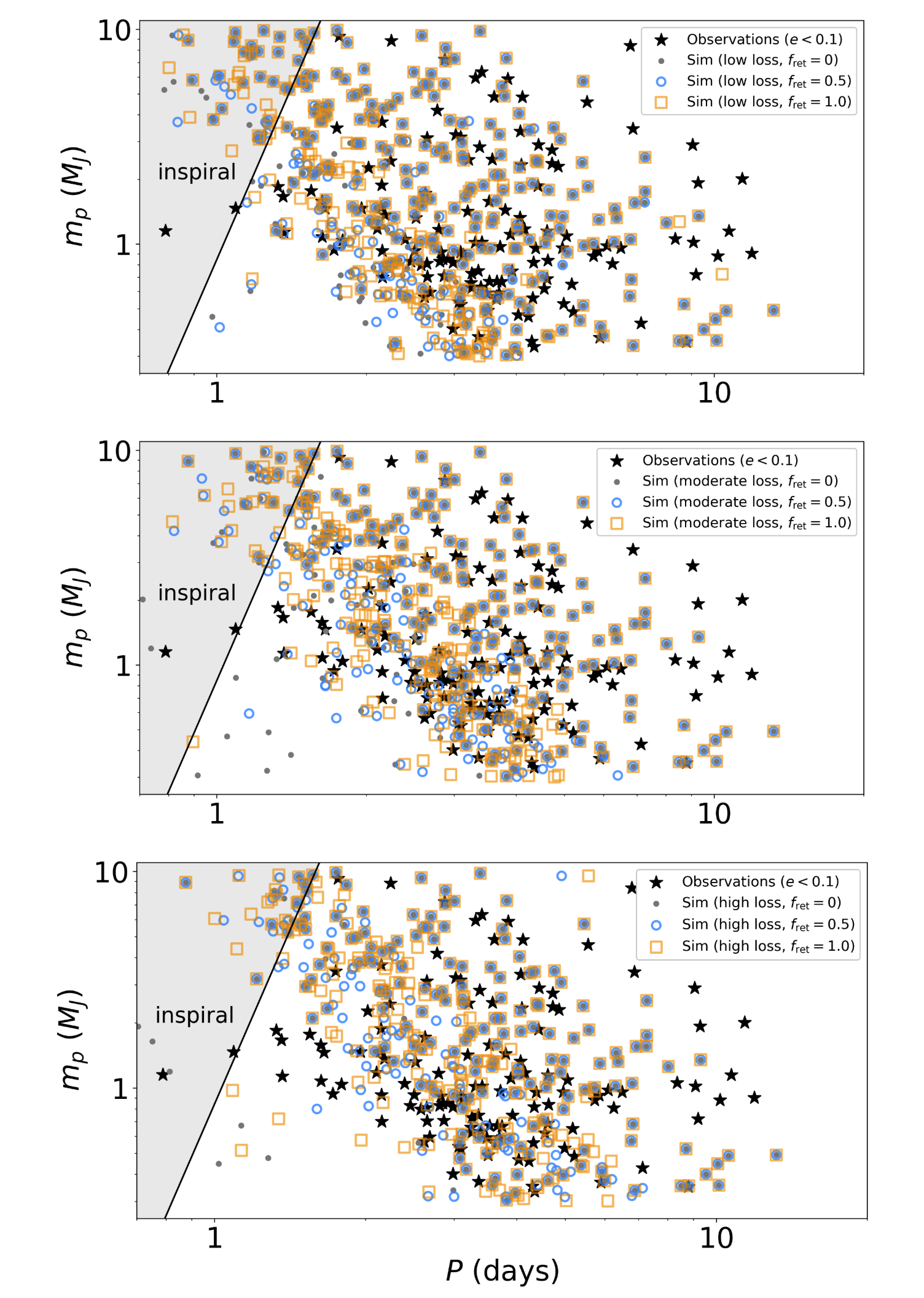}
\caption{\footnotesize Mass-period distribution for observed hot Jupiters (black stars) that are relatively circularized (measured eccentricities $e<0.1$), along with the final state of surviving hot Jupiters in the simulations (gray dots have $f_{\rm ret} = 0$, blue open circles have $f_{\rm ret} = 0.5$, and orange open squares have $f_{\rm ret} = 1$). The observed sample of hot Jupiters is taken from the NASA Exoplanet Archive (accessed August 22, 2025) \citep[e.g.,][]{Akeson+13,PSCompPars}. We select systems with $0.3M_J<m_p<10M_J$ and $a_p<0.1$ au. The top row corresponds to the low-loss case, the middle row corresponds to moderate loss, and the bottom row corresponds to high loss. The shaded region shows where inspiral and merger with the host star from stellar tidal dissipation is efficient, calculated using the calibration from \cite{Hansen10}.}
\label{fig:massperiod_sims}
\end{center}
\end{figure*}

\begin{figure*}[ht]
\begin{center}

\includegraphics[width=7in]{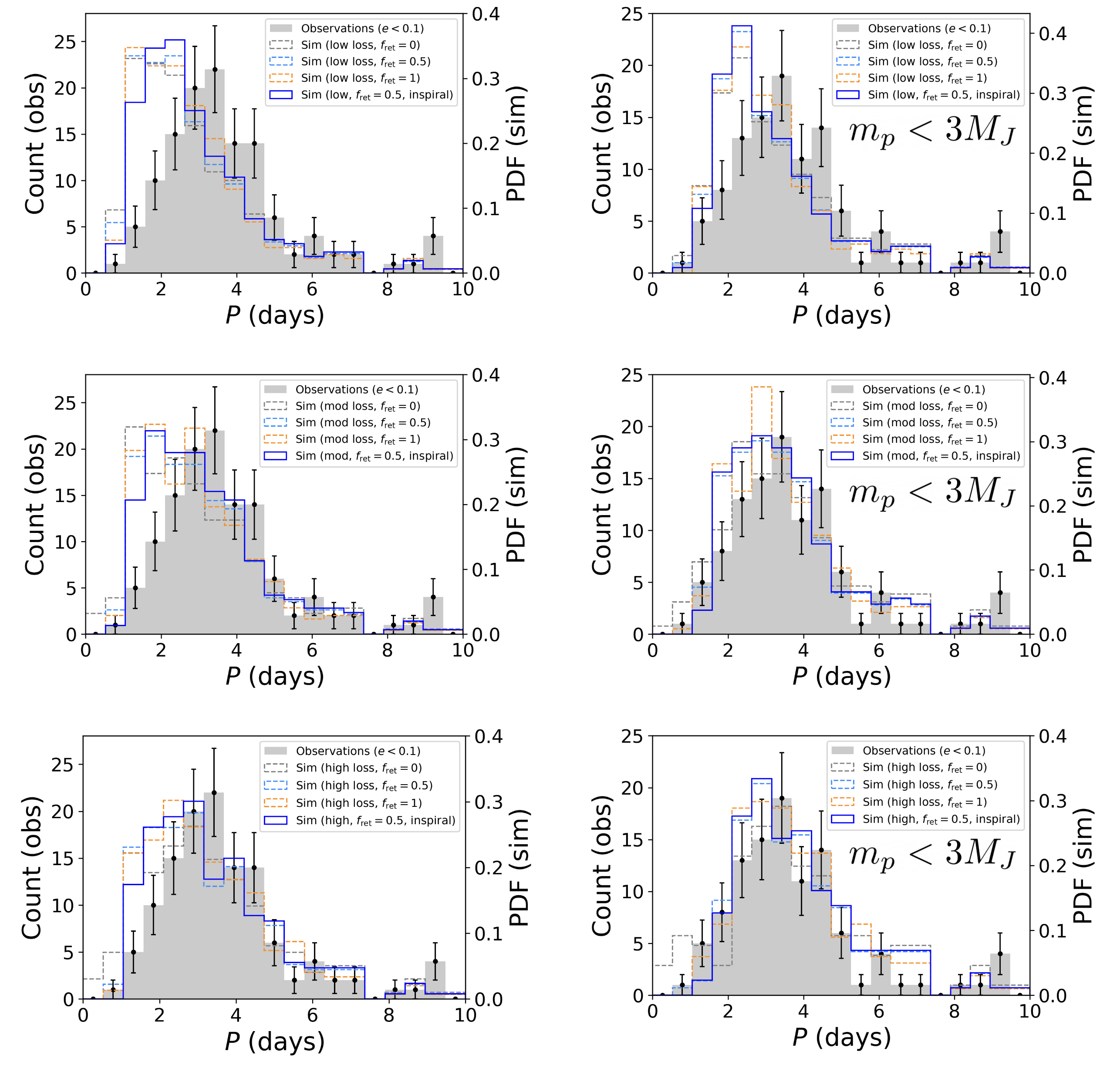}
\caption{\footnotesize Comparison between the observational period distribution (displayed in counts) and simulated period distribution (displayed as a PDF). The shaded gray histogram shows the period distribution of observed hot Jupiters with $e<0.1$. The period distributions for the simulated hot Jupiters are shown (dashed gray corresponds to $f_{\rm ret} = 0$, dashed blue to $f_{\rm ret} = 0.5$, and dashed orange to $f_{\rm ret} = 1$). The solid blue curves show the distributions with systems that undergo inspiral and merger in $\sim10^9$ yr removed, for the $f_{\rm ret} = 0.5$ case in each respective panel. The top row corresponds to the low loss case, the middle row to the moderate loss case, and the bottom row to the high loss case. The left column includes all planetary masses, while in the right we restrict to only planets with mass $<3 M_J$.
}
\label{fig:P_distributions}
\end{center}
\end{figure*}

We next investigate the population of hot Jupiters in mass-period space. In Figure \ref{fig:massperiod_sims}, we compare the simulated hot Jupiters with observations. Black stars show the observed sample. Gray dots correspond to $f_{\rm ret}=0$, blue open circles to $f_{\rm ret}=0.5$, and orange open squares to $f_{\rm ret}=1$. From top to bottom, the panels respectively show the low, moderate, and high mass loss cases.  The region where inspiral and merger from stellar tidal dissipation is shaded \citep[][]{Hansen10}.

Broadly, in both the simulations and observations, there is a cluster of planets between $\sim0.3-3 M_J$ and $\sim2-4$ days. The frequency of planets falls off at higher masses and periods. The lower left of each plot shows a relative dearth of planets, due to tidal disruption during highly eccentric passages. This threshold is explored in more analytical detail in Section~\ref{sec:analytical}. 

As the degree of mass loss increases from low to moderate to high, the population shifts to higher periods, owing to the stronger angular momentum return for higher mass loss fractions. Additionally, we can observe the effects of varying $f_{\rm ret}$. For $f_{\rm ret} = 0$, angular momentum return is unable to save many planets from disruption, revealing a lack of planets at low masses and periods. When comparing $f_{\rm ret} = 0.5$ and $f_{\rm ret} = 1$, we see that systems with $f_{\rm ret} = 1$ generally fall at slightly higher periods than those with $f_{\rm ret} = 0.5$, as expected given the higher angular momentum return fraction. This separation is especially noticeable at shorter periods, closer to the Roche limit. However, the similarity between these distributions can be attributed to the competition between mass loss fraction and angular momentum return. Though systems with $f_{\rm ret} = 1$ have a higher fraction of lost angular momentum returned than those with $f_{\rm ret} = 0.5$, these systems lose less mass over repeated passages, and thus there is less angular momentum to be returned.

Inward of the inspiral line, the observations reveal a relative dearth of planets. In our simulations, we stop the integration once a hot Jupiter forms before stellar tides are able to destroy the planet, leaving the planets in place in mass-period space. On long timescales, most of these planets are likely to be removed from the population, which will leave a void in mass-period space consistent with observations.

Outward of the inspiral line, we see that the simulations overpredict the number of planets at higher masses ($ \gtrsim 3M_J$) that survive as hot Jupiters. It is possible that the source population of cold Jupiters in nature is more bottom heavy than the $dn/dm_p \propto m_p^{-1}$ power law that was chosen in the simulation initial conditions. As discussed in Section~\ref{sec:ml_frac}, there is a discrepancy in the observed mass distribution for cold and hot Jupiters, which may be skewing the final results here. An additional possibility is that the overabundance of massive planets in the simulations reveals the limitations of the tidal models employed in this work. It is possible that the inspiral criterion may need modification at the high mass end, or that additional tidal effects that are not included in our models, such as tidal heating, may be needed to resolve this discrepancy. These possibilities warrant future study to better understand how mass loss affects the population of high mass super-Jupiters.

In Figure \ref{fig:P_distributions}, we show the period distributions of simulated and observed hot Jupiters, accounting for both stellar tides and the selection of planets in different mass ranges. The left column shows the full population of planets from $0.3M_J<m_p<10M_J$, and the right column shows the population of planets with $m_p<3M_J$ to avoid the discrepancy at high masses. The shaded gray histograms shows the period distribution of observed hot Jupiters with $e<0.1$. The probability density functions for the simulated hot Jupiters are overplotted (dashed gray corresponds to $f_{\rm ret} = 0$, dashed blue to $f_{\rm ret} = 0.5$, and dashed orange to $f_{\rm ret} = 1$). The solid blue curves show the distributions with systems that inspiral and merge with the star in $\sim10^9$ yr removed, for the fiducial $f_{\rm ret} = 0.5$ case in each respective panel. We note that the inspiral is calculated using the parameters for each individual system.

The observed distribution shows the well-known ``three-day pileup'' of hot Jupiters \citep[e.g.,][]{Gaudi+05,Wright+09,Yee+25}, which has previously been difficult to reconcile with EKL-driven high-eccentricity migration scenarios \citep[e.g.,][]{Dawson+13,Petrovich15b}. Indeed, for low fractions of mass loss and angular momentum return, we see that the simulations produce an excess of systems near $\sim2$ days, roughly twice the characteristic Roche limit. However, higher fractions of mass loss and angular momentum can push the peak of the population outwards to $\sim3$ days. This peak in the simulated period distribution becomes especially pronounced when tidal disruption via inspiral due to stellar tidal dissipation is accounted for using the calibration of \cite{Hansen10} (see Eq.~\ref{eq:hansen}), and the agreement appears especially strong in the sample that removes the highest mass planets.

To obtain a statistical comparison, we use a two-sample Kolmogorov-Smirnov (KS) test to assess the similarity between the observed sample and simulations. Taking the fiducial value of $f_{\rm ret} = 0.5$ for the full planet population and accounting for stellar tides, we obtain KS-test $p$-values of $\sim10^{-7}$ (low loss), $\sim10^{-3}$ (moderate loss), and $\sim0.07$ (high loss). For the population of planets below $3M_J$, we obtain $p$-values of $\sim10^{-3}$ (low loss), $\sim0.43$ (moderate loss), and $\sim0.80$ (high loss). These results indicate statistical agreement between the high loss case and the observations, and they suggest agreement for the moderate loss case when the highest mass planets are excluded.

Altogether, these results demonstrate the feasibility of EKL-induced high-eccentricity migration in producing a period distribution consistent with the observed distribution. Without mass loss or angular momentum return, planets tend to pile up near periods of $\sim2$ days. With moderate to high amounts of loss \citep[similar to or stronger than the first-passage results from][]{Guillochon+11}, enough angular momentum may be returned to push planets to pile up near $\sim3$ days. Stellar tides may also act to remove planets with periods shorter than 3 days, aiding in the construction of the pileup. Although these results are sensitive to the choice of planetary mass function, statistical agreement can be obtained for a typical power-law distribution.

\section{Discussion}
\label{sec:discussion}

In this work, we consider a model of a star+planet pair orbited by a distant stellar companion. We study the secular perturbations induced by the companion and the tidal mass loss experienced by the planet at high eccentricities. In population syntheses, we find that many giant planets may undergo partial tidal disruption and ultimately survive as stripped hot Jupiters. A sufficient fraction of planets survive high-eccentricity migration to produce agreement with observed hot Jupiter occurrence rates. In this section, we discuss additional physical processes that may contribute to the production or disruption of hot Jupiters via the high-eccentricity migration channel. We also discuss observational tests of these models.

\subsection{Additional factors in stellar multiples}

This work has focused on the production of hot Jupiters in stellar binaries. Giant planets tend to form more frequently around host stars with metallicities above the solar value \citep[e.g.,][]{Fischer+05}, though an anti-correlation has been observed with stellar metallicity and multiplicity \citep[e.g.,][]{Badenes+18}. Accounting for these trends may lower the production rate of hot Jupiters from stellar EKL. However, \cite{Fischer+05} observes a correlation between stellar metallicity and planetary multiplicity, and various high-eccentricity migration scenarios invoke planetary companions \citep[e.g.,][]{Ford+05,Wu+11,Teyssandier+19,Naoz+11,Petrovich+15a}. It is possible that planetary and stellar-induced migration operate in a complementary way around stars of varying metallicity to produce the population of hot Jupiters.

While some studies have found a correlation between the existence of hot Jupiters and the presence of a wide stellar companion \citep[e.g.,][]{Knutson+14,Ngo+15,Ngo+16,EelesNolle+25}, others have found a similar rate of hot Jupiters around single stars and in wide binaries \citep[e.g.,][]{Moe+21}. Those hot Jupiters around single stars could have formed via stellar EKL, and the companion was later unbound via white dwarf kicks \citep[][]{Stephan+24}. It is also possible that those in single star systems form completely unaided by a companion star and instead form via the aforementioned planet-planet migration pathways \citep[e.g.,][]{Ford+05,Wu+11,Teyssandier+19,Naoz+11,Petrovich+15a}. Indeed, our stellar EKL models produce occurrence rates of roughly $\sim0.5\%$. While this is consistent with transit-based estimates \citep[e.g.,][]{Petigura+18}, the dearth of planets relative to the $\sim1\%$ estimates of radial velocity surveys \citep[e.g.,][]{Wittenmeyer+20} suggests that additional mechanisms may be at play to account for the remaining hot Jupiter population.

There are additional factors that may boost the production of hot Jupiters in stellar multiples beyond the estimates in this work. We have focused on the production of, at most, one hot Jupiter per stellar binary system. Recent work shows that up to $\sim9\%$ of hot Jupiter systems may form a ``double'' system with one planet orbiting each star \citep[][]{Liu+25}. Accounting for this possibility would strengthen the agreement between the occurrence rate estimates and observations.

Additionally, while we have factored triple (and higher multiple) systems into the calculation of the occurrence rates, we have not accounted for the possibility of four-body interactions. A tertiary star may induce EKL oscillations in the secondary star, in turn leading to planetary eccentricity excitations and tidal interactions with the host star. This ``eccentricity cascade'' mechanism has been shown to potentially explain the architecture of the HAT-P-7 system \citep[][]{Yang+25}. Such an effect may also enhance the rate of hot Jupiter production beyond the estimates in this work.

\subsection{Planetary companions}
\label{sec:planetarycompanions}

This work has considered the case of single giant planets in stellar binaries, as prior studies have done \citep[e.g.,][]{Naoz+12,Petrovich15b,Weldon+25}. In reality, some systems may have additional giant planets \citep[e.g.,][]{Wu+23,Zink+23}. Mutual gravitational interactions between neighboring giant planets may quench the EKL effect from any stellar companions and lower the hot Jupiter production rate \citep[e.g.,][]{Denham+19,Wei+21}. However, the initial configurations of such systems are currently unclear; it is possible that some systems inhabit the regime where EKL quenching is weak. The aforementioned complementarity between planetary and stellar multiplicity around stars of varying metallicity may also prevent EKL quenching configurations from arising \citep[e.g.,][]{Fischer+05}. Future observational data may shed light on the prevalence and nature of multi-planet systems with stellar companions.

In some systems, planetary companions may themselves induce the EKL effect, causing the inner planet to become a hot Jupiter \citep[e.g.,][]{Naoz+11,Petrovich+15a,Petrovich+16}. Though the initial conditions for such systems are also poorly constrained, previous simulations that have attempted to estimate the rate of hot Jupiter production from planet-planet interactions have found that many are completely tidally disrupted \citep[e.g.,][]{Petrovich+15a}. Because previous studies employed a hard boundary for tidal disruption, the treatment in this study may also boost the survival of planets perturbed by companion planets. Therefore, applications of these models to other high-eccentricity migration scenarios may broadly alleviate a key discrepancy between theory and observation.

\subsection{Applications to other high-eccentricity migration scenarios}

We have studied the stellar EKL mechanism as a test case due to the relatively well-constrained initial conditions of stellar companions. The treatment of mass loss and angular momentum return in this work may be applied to other high-eccentricity migration scenarios. Here, the relaxation of the tidal disruption criterion compared to prior studies is justified, in part, because differential secular precession acts to wind planetary debris away from stripped planetary remnants, making planets less likely to be disrupted than calculated in \cite{Guillochon+11} (as described in Section \ref{sec:tidalmassloss}). However, for systems that remain on unperturbed Keplerian orbits (e.g., in the case of planet-planet scattering), this assumption may not hold, and planets may be disrupted more frequently due to increased re-accretion of debris. In cases where the assumption of minimal encounters between remnants and debris is valid, the fraction of planets nearing the Roche limit depends on the efficiency of excitation from a population of perturbers. Therefore, the factor of enhancement of hot Jupiter production will depend on the details of the chosen mechanism, warranting future study for the secular chaos mechanism \citep[e.g.,][]{Wu+11,Teyssandier+19} and planet-planet EKL in either coplanar \citep[e.g.,][]{Petrovich+15a} or inclined \citep[e.g.,][]{Naoz+11} configurations. 

The mass loss at pericenter in our work occurs over $\mathcal{O}(10^2-10^4)$ years, which is shorter than the characteristic Myr-Gyr timescales over which secular mechanisms tend to act \citep[e.g.,][]{Wu+11,Naoz+11,Petrovich+15a}. Indeed, phases of very high eccentricity ($e\sim0.99$) represent a small fraction of the evolution. Therefore, the formalism presented here can be used to assess the response of planets and their orbits to mass loss during episodes of high eccentricity. As a result, the trends in the mass dependence of surviving and disrupted planets should hold for other formation channels; however, the exact mass distributions may depend on the chosen dynamical mechanism. The return of angular momentum to the orbit is also expected to occur in a manner similar to the stellar EKL case, potentially pushing the predicted period distributions outward.

\subsection{Dynamical tides and tidal heating}

Previous population synthesis studies have included the effect of dynamical tides on giant planet migration \citep[e.g.,][]{Vick+19}. They show that oscillatory modes excited in the planet can grow chaotically over multiple orbits and dissipate non-linearly. As a consequence, rapidly decreasing $e_p$ and increasing $q$ can save planets from tidal disruption. At the population level, \cite{Vick+19} finds that the hot Jupiter production rate is increased by a factor of $\sim1.6$ from this effect. A detailed study combining dynamical tides with the mass loss prescription in this work may yield an even greater fraction of surviving hot Jupiters.

Recently, \cite{Yu+25} showed that the dynamical tide leads planets to begin losing mass near $q\sim2.7r_t$. While this threshold agrees with \cite{Guillochon+11} for the same orbital eccentricity ($e_p=0.9$), the hot Jupiters in our study tend to originate from even more eccentric orbits ($e_p\sim0.99$). As discussed in \cite{Yu+25}, more eccentric orbits spend less time near periapse, requiring more passages for the dynamical tide perturbation to build before significant mass loss occurs. On the other hand, the dynamical tide may be more easily excited in these systems, shifting the mass loss threshold outward. While future studies may explore the competition between these effects in greater detail, we expect mass loss to remain a gradual function of periapse distance, rather than set by a hard boundary. 

Our study has focused on the survival of cold Jupiters through passages of extreme eccentricity, a phase that, under previous treatments, destroys a substantial fraction of planets. We have shown that a large fraction of these disrupted planets may survive as stripped hot Jupiters. It is possible that tidal heating during the intermediate warm Jupiter phase may lead to further episodes of mass loss and potential tidal disruption \citep[e.g.,][]{Rozner+22,Glanz+22,Yu+24,Lu+25,Gao+25,Hallatt+25b}. While at present it remains unclear exactly how tidal heating is deposited in giant planets, the impacts of various tidal heating models on the population will be explored in future work. The dynamical outcomes obtained here may inform the initial conditions in the later phases of the evolution where tidal heating becomes important.

\subsection{Observational tests}

Observational probes of planetary tidal mass loss are underway. Recently, \cite{Saidel+25} conducted a survey of atmospheric escape from hot Jupiters orbiting F-type stars. They  found that planets with higher Roche lobe filling factors display stronger levels of mass loss, for Roche lobe filling factors between $\sim 0.2-0.55$, a trend consistent with our expectations. Future surveys may study planets undergoing more extreme levels of mass loss. An infrared transient from tidal engulfment of a giant planet has already been observed \citep[][]{De+23}, and future transient searches may probe a range of tidal encounter scenarios. Additionally, the detection of infrared excess from accretion structures in systems hosting giant planets may indirectly shed light on mass loss \citep[as studied by][for the Roche lobe overflow of giant planets induced by stellar tides]{Hallatt+25}. They find that the degree of infrared excess may constrain the fraction of mass retained by the system, providing tests of the angular momentum return models in this work. 

Our framework predicts that for stellar companion projected separations below $\sim300$ au, hot Jupiters that lose mass dominate the population (see Figure \ref{fig:analytic_model}). Therefore, while evidence for mass loss may be found in systems with more distant companions, it should be more likely in systems with close-in companions. We also show in Figure \ref{fig:analytic_model} that the observed distribution of stellar companion separations to hot Jupiters is in agreement with the model prediction in this work, though we note that the sample is incomplete and heterogeneous. Upcoming surveys may provide additional constraints on the nature of giant-planet-hosting binaries. GAIA DR4 may provide a wealth of information on the architectures of such systems for more robust comparisons with models, particularly in the super-Jupiter regime ($m_p>3M_J$) \citep[e.g.,][]{Lammers+26}.

An additional prediction of the model is that stripped giant planets may appear younger than their host stars, due to tidal mass loss, on average, increasing the planetary radius relative to an undisturbed evolution. More robust constraints on the ages of planetary systems may test this prediction. In particular, studying the ages of systems harboring highly eccentric cold Jupiters or distant warm Jupiters may provide a test of this effect, as close-in warm and hot Jupiters may be subject to additional heating that alters the planetary radius.

\section{Conclusions}
\label{sec:conclusion}

We perform analytical and numerical population synthesis studies of giant planets to determine the efficiency of high-eccentricity migration in producing hot Jupiters. As a test case, we explore the generation of high eccentricities via the stellar EKL mechanism and examine planetary survival against tidal disruption at high eccentricities. Other works have suggested that relaxing the criterion for tidal disruption may boost the rate of production \citep[e.g.,][]{Petrovich15b,Yu+24}; however, this work is the first to explore a realistic treatment of the planetary response to eccentric mass loss for the population of hot Jupiters. Our main conclusions are:

\begin{itemize}
    \item The hot Jupiter formation rate from stellar EKL may be increased by a factor of $\sim2-3$ when accounting for tidal mass loss and angular momentum return, in comparison to prior population studies \citep[e.g.,][]{Naoz+12,Petrovich15b,Anderson+16} that set a hard boundary for tidal disruption.

    \item The hot Jupiter occurrence rate around FGK stars predicted by the models is $\gtrsim0.5\%$ (see Table \ref{tab:sims} and Figure \ref{fig:rates}), which is entirely consistent with transit-based estimates \citep[e.g.,][]{Howard+12,Fressin+13,Petigura+18} and can account for the majority of hot Jupiters in radial velocity based estimates \citep[e.g.,][]{Mayor+11,Wright+12,Wittenmeyer+20}.

    \item Angular momentum return from mass accreted onto the star can lead hot Jupiters to pileup at $\sim3$ day orbital periods, a well-known feature in the observed population \citep[e.g.,][]{Gaudi+05,Wright+09}. The overall simulated period distribution statistically agrees with observations (see Figure \ref{fig:P_distributions}).

    \item The amount of mass loss necessary for the model to produce a match to observations is consistent with hydrodynamical simulations by \cite{Guillochon+11}, for the first passages of a planet near the host star, prior to total disruption. Lower levels of mass loss increase the rate of hot Jupiter production, but do not lead to the three-day pileup.

   \item The overall hot Jupiter rate depends on, but is not highly sensitive to, the initial locations of planets or companion stars for typical choices (see Figure \ref{fig:analytic_model}). While stronger average perturbations lead to lower periastra and enhanced tidal migration, populations with weaker perturbers generally have fewer planets being tidally disrupted, leading more to survive as stripped hot Jupiters.

   \item The observed distribution of hot Jupiter stellar companion separations is in agreement with the analytical model at the level of $\sim1-2$$\sigma$ (see panel (d) of Figure \ref{fig:analytic_model}). Both reveal a deficit of close-in companions ($\lesssim300$ au) relative to field stars.

    \item Eccentric mass loss leads to an average mass reduction of $\sim5\%$, though individual systems may experience significantly higher loss. However, low mass planets tend to be pushed out of the hot Jupiter population (i.e., fall below $0.3M_J$) and higher mass planets are better at retaining their mass. As a consequence, the average mass of the population of surviving hot Jupiters is predicted to be $\sim5-10\%$ more massive than the average mass of the population of cold Jupiters.

    \item The majority of hot Jupiters in the models are planets that experience some degree of tidal stripping. Therefore, we predict that the majority of hot Jupiters in nature are stripped systems. Stripped giant planets may generally appear younger than their host stars, due to tidal mass loss altering the planetary radius relative to an undisturbed evolution. 
\end{itemize}

Many lines of observational evidence suggest that high-eccentricity migration plays a significant role in hot Jupiter formation, including orbital misalignments in the hot and warm Jupiter populations \citep[e.g.,][]{Triaud+10,Albrecht+12,Rice+22}, high stellar companionship around hot Jupiters \citep[e.g.,][]{Knutson+14,Ngo+15,Ngo+16}, and the high eccentricities of potential source cold Jupiters \citep[e.g.,][]{Winn+15,Weldon+25}. Despite this evidence, a long-standing issue has been that the theoretical production rate of hot Jupiters is far too low \citep[e.g.,][]{Naoz+12,Petrovich15b,Anderson+16}. We show that accounting for mass loss and angular momentum return may save many planets from tidal disruption. The significantly higher survival rate that we find provides theoretical support for high-eccentricity migration being a dominant driver of hot Jupiter production.
\\
\\
The authors thank the anonymous referee for helpful comments. The authors also thank Hang Yu, Tim Hallatt, Sarah Millholland, and Gongjie Li for useful discussions. The authors acknowledge the support of NASA XRP grant 80NSSC23K0262. S. N. thanks Howard and Astrid Preston for their generous support. The authors also acknowledge the use of the UCLA cluster \textit{Hoffman2} for computational resources. This research has made use of NASA’s Astrophysics Data System Bibliographic Services. This research has also made use of the NASA Exoplanet Archive, which is operated by the California Institute of Technology, under contract with the National Aeronautics and Space Administration under the Exoplanet Exploration Program. This research has also made use of data obtained from or tools provided by the portal exoplanet.eu of The Extrasolar
Planets Encyclopaedia.

\software{
    NumPy \citep{numpy},
    SciPy \citep{scipy},
    Matplotlib \citep{matplotlib},
    Mathematica \citep{Mathematica},
    WebPlotDigitizer \citep{WebPlotDigitizer}
}

\newpage

\bibliography{paperbib, softwarebib}{}
\bibliographystyle{aasjournal}

\appendix

\section{MESA Grids}
\label{app:mesa_cooling}

\begin{figure*}[ht]
\begin{center}

\includegraphics[width=4in]{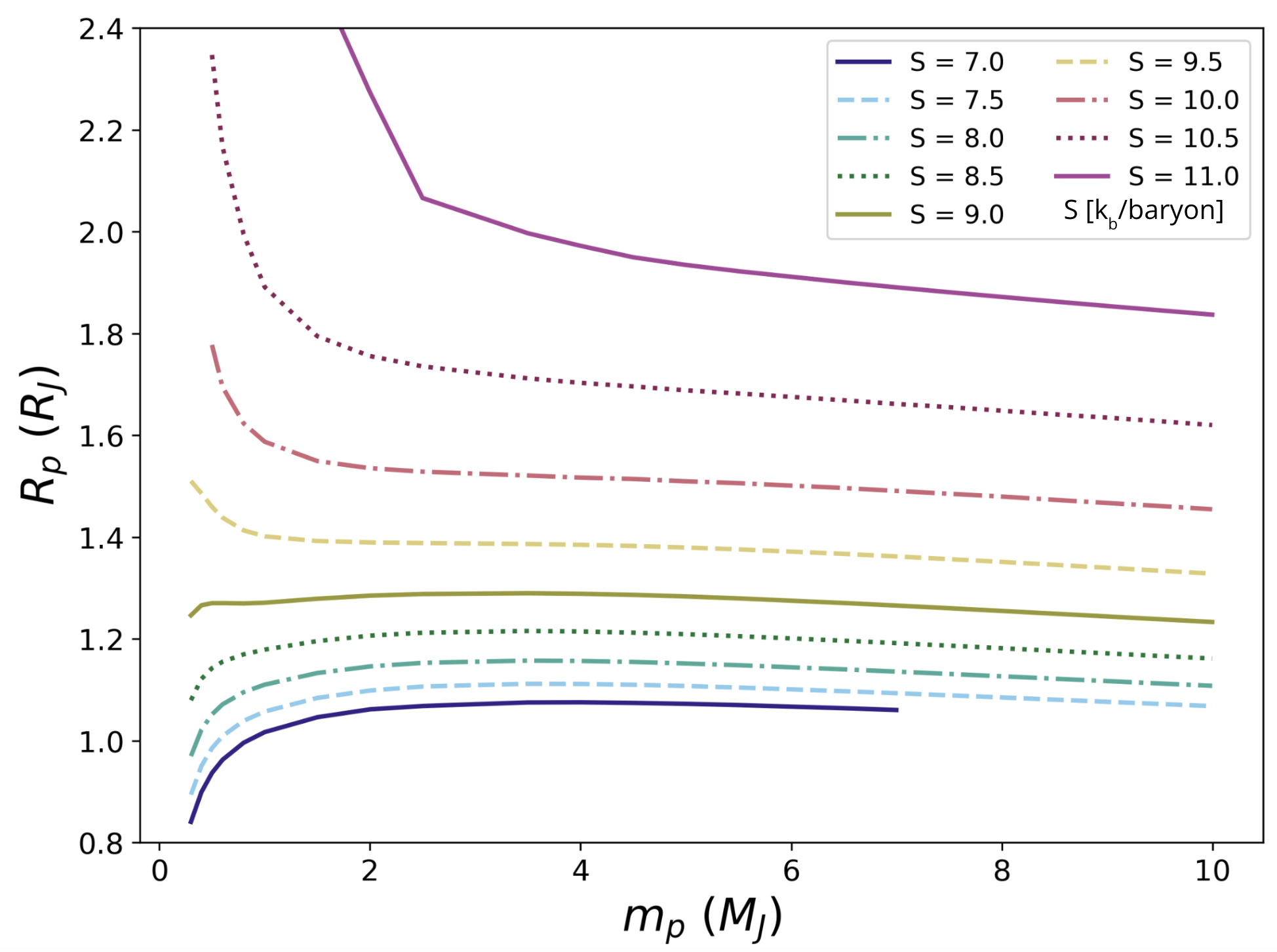}
\caption{\footnotesize Curves of constant central entropy showing planetary radius as a function of planetary mass from MESA.}
\label{fig:mesa_grids}
\end{center}
\end{figure*}

In Figure \ref{fig:mesa_grids}, we show mass-radius relations at constant central entropy obtained in MESA. Given the planet mass and entropy at some time, we linearly interpolate along the grid to model the radius. A planet that loses mass will adjust its radius on a dynamical timescale \citep[e.g.,][]{Paczynski+72}. This timescale is shorter than the time for significant cooling to occur, and giant planets are almost fully convective and hence isentropic throughout \citep[e.g.,][]{Arras+06}. As a planet loses mass, it will adjust its radius to that of a planet with the changed mass but the same internal entropy.

\begin{figure*}[ht]
\begin{center}

\includegraphics[width=6in]{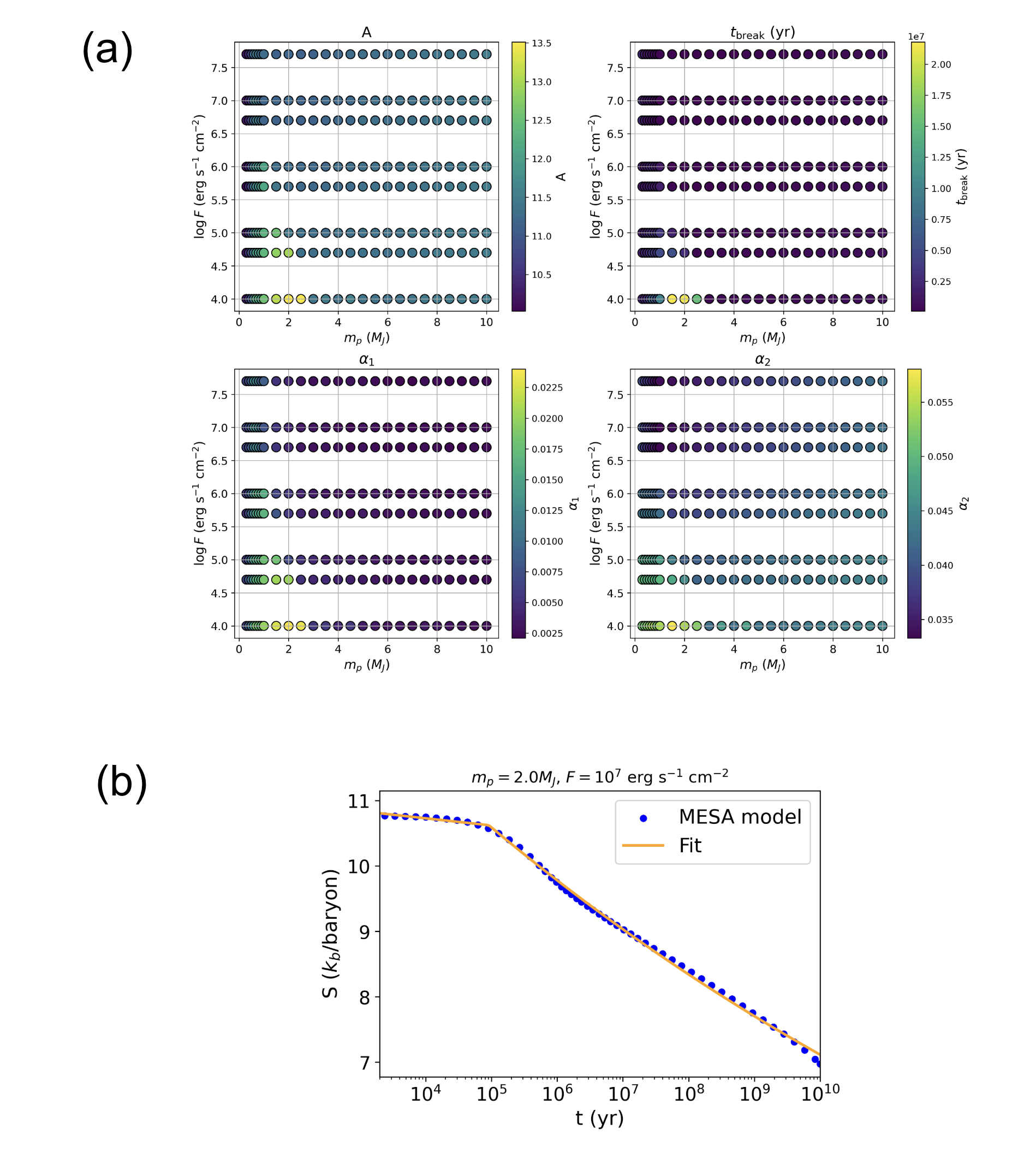}
\caption{\footnotesize Panel (a): The fitted parameters in Eq.~(\ref{eq:S_fit}) from MESA evolutionary models. The color code shows $A$ (top left), $t_{\rm break}$ (top right), $\alpha_1$ (bottom left), and $\alpha_2$ (bottom right) as a function of the logarithm of stellar flux and the planetary mass. Panel (b): Example fit to a MESA cooling curve using the fit parameters from panel (a) for a $2.0M_J$ planet with stellar flux of $10^7$ erg s$^{-1}$ cm$^{-2}$.}
\label{fig:mesa_fits}
\end{center}
\end{figure*}

To model planet cooling, we fit entropy as a function of time to each of our MESA models using a broken power law
\begin{equation}
    S(t) =
\begin{cases}
A t^{-\alpha_1}, & t < t_{\mathrm{break}} \\
A t_{\mathrm{break}}^{\alpha_2 - \alpha_1} t^{-\alpha_2}, &  t \geq t_{\mathrm{break}}
\end{cases}
\label{eq:S_fit}
\end{equation}
where we show the values for $A$, $t_{\rm break}$, $\alpha_1$, and $\alpha_2$ as a function of planet mass and incident stellar flux in panel (a) of Figure \ref{fig:mesa_fits}. We linearly interpolate along this grid to obtain fit parameters for planet masses and stellar fluxes between the sampled values. We manually verified that each fit produces a close match to the MESA data.  An example fit is shown in panel (b) of Figure \ref{fig:mesa_fits}.

\section{Stellar Obliquity Distribution}
\label{app:obliquities}

\begin{figure*}[ht]
\begin{center}

\includegraphics[width=7in]{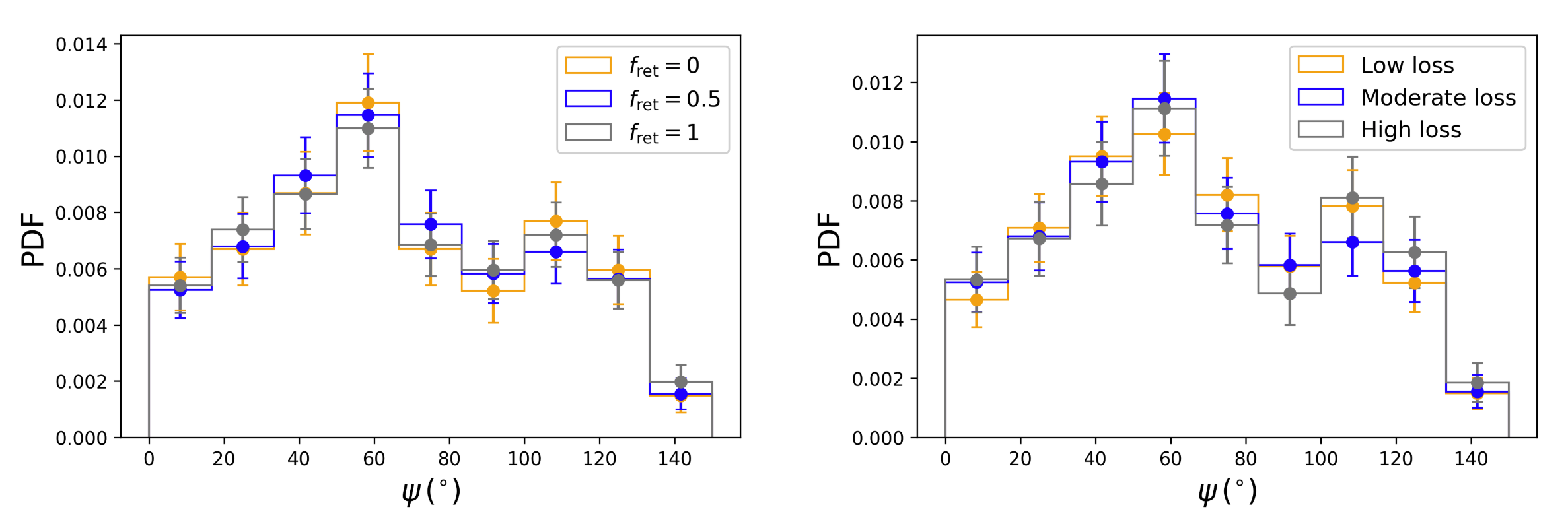}
\caption{\footnotesize Left: Distributions of stellar obliquities $\psi$ for hot Jupiters in the moderate-loss numerical simulations. The orange curve corresponds to $f_{\rm ret}=0$, the blue curve to $f_{\rm ret}=0.5$, and the gray curve to $f_{\rm ret}=1$. Right: Distributions of obliquities for hot Jupiters in the simulations with $f_{\rm ret}=0.5$. The orange curve corresponds to the low loss prescription, the blue curve to moderate loss, and the gray curve to high loss.}
\label{fig:obliquities}
\end{center}
\end{figure*}

Here, we examine the distribution of final stellar obliquities $\psi$ of hot Jupiters in the numerical simulation runs. In the left panel of Figure \ref{fig:obliquities}, we show the distributions in the moderate loss runs with varied $f_{\rm ret}$. In the right panel, we show the distributions in the $f_{\rm ret}=0.5$ case while varying the mass loss prescription. The bimodal distribution that we find is consistent with prior studies \citep[e.g.,][]{Naoz+12,Petrovich15b,Anderson+16}. There are no statistically significant differences between the distributions for varying $f_{\rm ret}$ and mass loss prescription. Therefore, we conclude that for the levels of mass loss studied in this work, mass loss has a negligible effect on the final stellar obliquity distribution. More catastrophic loss may alter the stellar obliquity distribution in a more significant way, though such studies are beyond the scope of this work. 

\end{document}